\documentclass[prd,aps,preprint,tightenlines,superscriptaddress,nofootinbib,showpacs,showkeys]{revtex4}
\usepackage{float}
\usepackage{mathrsfs}
\usepackage{amsfonts}
\usepackage{array}
\usepackage{epsfig}
\usepackage{amsmath}    % need for subequations
\usepackage{amssymb}   % defines \lesssim, etc
\usepackage{graphicx}   % need for figures
\usepackage{verbatim}   % useful for program listings
\usepackage{color}      % use if color is used in text
\usepackage{subfigure}  % use for side-by-side figures
\usepackage{multirow}

\begin{document}

\title{
Higgs Boson Cross Section from CTEQ-TEA Global Analysis}

\author{Sayipjamal Dulat}
\email{sdulat@msu.edu}
\affiliation{
School of Physics Science and Technology, Xinjiang University,\\
 Urumqi, Xinjiang 830046 China }
\affiliation{
Department of Physics and Astronomy, Michigan State University,\\
 East Lansing, MI 48824 U.S.A. }
\author{Tie-Jiun Hou}
\affiliation{
Institute of Physics, Academia Sinica, Taipei, Taiwan 115 }
\author{Jun Gao}
\affiliation{
Department of Physics, Southern Methodist University,\\
 Dallas, TX 75275-0181, U.S.A. }
\author{Joey Huston}
\affiliation{
Department of Physics and Astronomy, Michigan State University,\\
 East Lansing, MI 48824 U.S.A. }
\author{Pavel Nadolsky}
\affiliation{
Department of Physics, Southern Methodist University,\\
 Dallas, TX 75275-0181, U.S.A. }
\author{Jon Pumplin}
\affiliation{
Department of Physics and Astronomy, Michigan State University,\\
 East Lansing, MI 48824 U.S.A. }
\author{Carl Schmidt}
\affiliation{
Department of Physics and Astronomy, Michigan State University,\\
 East Lansing, MI 48824 U.S.A. }
\author{Daniel Stump}
\affiliation{
Department of Physics and Astronomy, Michigan State University,\\
 East Lansing, MI 48824 U.S.A. }
\author{ C.--P. Yuan}
\email{yuan@pa.msu.edu}
\affiliation{
Department of Physics and Astronomy, Michigan State University,\\
 East Lansing, MI 48824 U.S.A. }

\begin{abstract}
We study the uncertainties of the Higgs boson production cross section
through the gluon fusion subprocess at the
LHC (with $\sqrt s=7, 8$ and $14$ TeV) arising from the uncertainties
of the parton distribution functions (PDFs) and of the value of the
strong coupling constant $\alpha_s(M_Z)$.
These uncertainties are computed by two complementary approaches,
based on the Hessian and the Lagrange multiplier methods
within the CTEQ-TEA global analysis framework.
We find that their predictions for the Higgs boson cross section are
in good agreement. Furthermore, the result of the Lagrange multiplier method
supports the prescriptions we have previously provided for using the Hessian method to
calculate the combined PDF and $\alpha_s$ uncertainties, and
to estimate the uncertainties at the $68\%$ confidence level by scaling them from the
$90\%$ confidence level.

\end{abstract}



\maketitle

\section{Introduction}

The data accumulated by the ATLAS and CMS experiments in Run 1 at the
Large Hadron Collider (LHC), at 7 and 8 TeV,
have allowed the study of Higgs boson production to move from the discovery phase to the
beginning of the precision measurement phase.
With the increased statistics of the data comes the need for a better understanding
of the theoretical uncertainties on the prediction of the cross section for Higgs boson production.
The production cross section, $\sigma_H$, of the primary subprocess, gluon-gluon ($gg$) fusion,
is known to next-to-next-to-leading order (NNLO) in perturbative QCD~\cite{Harlander:2000mg,
Catani:2001ic,Harlander:2001is,Harlander:2002wh,
Anastasiou:2002yz,Ravindran:2003um,Blumlein:2005im} in the infinite top mass limit
and to next-to-leading order (NLO) in electroweak corrections~\cite{Djouadi:1994ge,Aglietti:2006yd,
Degrassi:2004mx,Actis:2008ts,Actis:2008ug}.
Resummation predictions have improved the theoretical precision to next-to-next-to-leading order
plus next-to-next-to-leading logs (NNLO+NNLL)~\cite{Catani:2003zt}.
In addition, approximations of the next-to-next-to-next-to-leading order (NNNLO) QCD corrections
are available \cite{Moch:2005ky, Anastasiou:4379, Ball:2013bra}, and
corrections for finite $m_{\rm top}$ have been calculated \cite{Marzani:2544, Harlander:2997,
Harlander:3420, Harlander:2104, Pak:2998, Pak:4662}.
The theoretical uncertainties from these calculations of the hard cross section, which are still sizable due to the truncation of the perturbative series,
can be estimated by varying the renormalization and factorization
scales---traditionally by a factor of two around a central scale of Higgs boson mass $M_{H}$ or $M_{H}/2$.

Another sizable theoretical uncertainty for the $gg$ channel is that due to the uncertainty on the gluon parton distribution function in the relevant kinematic region.
The $gg$ PDF luminosity has been well-studied:
see, for example, the recent benchmark paper \cite{benchmark}.
There is reasonable, though not totally satisfying, agreement for the $gg$ PDF luminosity predictions
from the three global PDF fitting groups CTEQ-TEA (CT)~\cite{ct10nnlo},
MSTW \cite{mstw08}, and NNPDF~\cite{nnpdf23}.

Because of the importance of Higgs boson production, particularly in the $gg$ channel, it is important to re-examine the PDF uncertainties, as well as the related $\alpha_s$ uncertainty,
for the cross section $\sigma_{H}$.
In this paper, we calculate the PDF error $\delta\sigma_{H}$ using two different methods:
the well-known Hessian method~\cite{Pumplin:2001ct} and the Lagrange multiplier method~\cite{Stump:2001gu}.
We then compare the uncertainty determinations from these two methods.
This work follows the global analysis framework of the recently published CT10 NNLO PDFs~\cite{ct10nnlo} (but with some LHC data added in the study)
to estimate the error on $\sigma_{H}$ at center-of-mass energies 7, 8 and 14 TeV.

We are partly motivated in this paper by the results of the benchmark calculations in Ref.\ \cite{benchmark}.
Different PDFs give somewhat different predictions for the cross section $\sigma_{H}$,
and also somewhat different uncertainties for the predictions; cf. Appendix B.
Most of the benchmark calculations relied on error PDFs obtained using the Hessian method.
The central predictions are fairly consistent within the corresponding uncertainties.
But the result raises the question whether the Hessian method is sufficient.
We will therefore compare Hessian and Lagrange multiplier calculations to test
whether the Hessian method is trustworthy.

Our other goal is to examine how the PDF dependence in various Higgs
production channels is related through their shared degrees of
freedom, and which experiments in the CT10 analysis constrain the
gluon PDF in the kinematical region of Standard Model (SM) Higgs production. The two
analysis methods provide complementary quantitative information in this
regard.  Using the Hessian method, we observe the
pull of various error PDFs on the Higgs boson cross section measurement.
We also compute the correlation cosines~\cite{Nadolsky:2008zw}
between the gluon PDF and Higgs boson production cross sections in
$gg$ and  vector boson fusion (VBF) processes.
Within the context of the Lagrange multiplier method,
we identify the experimental data sets included in our global analysis
that correlate most strongly with the value of $\sigma_H$ via $gg$ channel.
This is done by introducing an equivalent Gaussian variable $S$
\cite{Lai:2010vv,Dulat:2013hea} for every fitted data set, which
provides an alternative to the usual chi-square distribution and
streamlines comparisons of constraints from heterogenous experiments.

In this study the calculation of the $gg\rightarrow{H}$ cross section
that we use is NNLO,
with the corrections obtained in the infinite $m_{\rm top}$ limit.  We
do not include finite
$m_{\rm top}$ corrections or approximations that go beyond NNLO.
These corrections can be calculated using the central PDFs, while their
 effect on the PDF errors would be small, of order 5\% of the corrections themselves.
Because our interest is in the PDF uncertainty for $\sigma_{H}$,
we omit these corrections.  By using a strictly NNLO calculation, we use the
 same order of approximation as the benchmark calculations in Ref.\ \cite{benchmark}.

The paper proceeds as follows.
In Section II, we discuss the theoretical background for the PDF global analysis and list the experimental inputs.
We then review the two frameworks for calculating PDF uncertainties:
the Hessian method and the Lagrange multiplier (LM) method.
The LM method is safer and more powerful;
but it requires the full global PDF analysis machinery to make predictions.
In Section III, we study the PDF uncertainties for Higgs boson production
using each of the two methods, and compare their results.
We also investigate the combined uncertainties in Higgs boson production coming from the PDFs and from $\alpha_s(M_Z)$ in the two methods.
In the Lagrange multiplier method we examine the correlation between $\sigma_{H}$ and $\alpha_s(M_Z)$
by constructing contour plots of the global $\chi^{2}$ as a function of
$\sigma_H$ and $\alpha_s(M_Z)$.
This result is then compared to that given by the Hessian method.
In Section IV, we investigate in more detail the correlation
between the Higgs boson production cross section and the PDFs, and the
origins of experimental constraints on the PDF dependence of the
$gg\rightarrow H$ cross section.
Finally, we compare the best-fit gluon PDFs, which correspond to various predictions for $\sigma_H$, to the error band of the CT10 NNLO PDFs.
Section V contains our conclusions.

\section{Global Analysis framework and uncertainty estimation}
\label{sec:GlobalAnalysis}

The CTEQ-TEA (CT) global analysis procedure has been extensively
discussed in previous papers~\cite{ct10nnlo}.
Here we review some aspects that are especially important for the
application to $\sigma_H$.

The CT parton distributions are obtained from global analysis of short-distance processes using a ``best-fit'' paradigm in which the PDFs are constructed by minimizing a global $\chi^2$ function.
The basic global chi-square function is defined by
\begin{eqnarray}
\chi^2 &=& \sum_{e} \chi_{e}^{2}(a,r),\label{eq:chi2}\\
\mbox{where \ \ \ }
\chi_{e}^{2}(a,r) &=& \sum_{\nu}
 \frac{\left[D_{\nu}-\sum_{k}r_{k}\beta_{k\nu} -
T_{\nu}(a)\right]^{2}} {\alpha_{\nu}^{2}} +\sum_{k}r_{k}^{2}.
\end{eqnarray}
Here $e$ labels an experimental data set and $\nu$ labels a data point in that data set.
$D_{\nu}$ is the central data value,
$\alpha_{\nu}$ is the uncorrelated error,
and $\beta_{k\nu}$ is the $k$-th correlated systematic error estimate.
These numbers are provided by the experimental collaborations.
$T_{\nu}(a)$ is the theoretical prediction, which is a function of a set of $n$ PDF
parameters, $\{a_{1}, \dots , a_{n}\}$.
In addition, $\{r_{k}\}$ is a set of Gaussian random variables;
thus, $r_{k}\beta_{k\nu}$ is a (correlated) shift applied to $D_{\nu}$ to represent the $k$-th systematic error.
We minimize the function $\chi^{{2}}(a,r)$
with respect to both the PDF parameters $\{a\}$ and the systematic shift variables $\{r_{k}\}$.
The result yields both the standard PDF model with parameters
$\{a_{0}\}$, and the optimal shifts $\{\widehat{r}_{k}\}$ to bring theory and data into agreement.
This minimum of $\chi^{{2}}$ represents the central fit to the data~\cite{ct10nnlo}.

Table~\ref{tab:EXP_bin_ID} shows the experimental data sets
employed in the current study.
Most of these data are the same as those used to produce the
CT10 NNLO PDFs~\cite{ct10nnlo}.
One of the new data sets included in this study is
the LHC data on $W^\pm$ and $Z$ production at ATLAS, labelled ID number 268,
which is a combined data set with the measurements of lepton rapidity
distributions from the $W^+$, $W^-$, and $Z$
boson productions and the charged lepton rapidity asymmetry~\cite{Aad:2011dm}
 at the LHC with 7 TeV center-of-mass energy.
This combined data set is analyzed with the full set of correlated
systematic errors, including the collider luminosity error, implemented.
Another new data set included in this study is the ATLAS single inclusive jet
production~\cite{Aad:2011fc} measured in anti-$k_T$ algorithm
with jet size parameter $R=0.6$, at the
LHC with a 7 TeV center-of-mass energy.
The ID number of this data set is 535.
We include these data because they come from LHC energies and may therefore be
particularly relevant to $\sigma_{H}$ at the LHC.

In this study, we use a somewhat more flexible parametrization
for the gluon distribution function than was used for the
CT10 NNLO PDFs.
Our interest is the process $gg\rightarrow{H}$,
which obviously depends directly on the gluon PDF, so
allowing a very flexible parametrization is important to reveal
the full range of the $\sigma_{H}$ uncertainty.

We will refer to the PDFs obtained in the current study
as CT10H NNLO PDF sets, to distinguish them from the standard CT10 NNLO PDFs.
The two global fits may differ slightly because (i) the CT10H fit includes
some LHC ($W$, $Z$ and jet) data not used for CT10; and (ii) the gluon distribution of CT10H
has a more flexible form.
However, the central fits for CT10H and CT10 differ little.
For that reason, and because CT10 NNLO PDFs have already
been used by LHC experiments,
we do not advocate that CT10H PDFs be used for general purpose PDFs.
CT10 NNLO will continue to be our standard general purpose PDFs,
until they are superseded by a successor to CT10 that includes input
from more extensive LHC data and the final results from HERA.

Table \ref{tab:EXP_bin_ID} indicates that the CT10H NNLO PDFs are in satisfactory agreement
with all the data sets included in the current analysis,
similar to the agreement seen in the CT10 NNLO analysis.
The relationship between the goodness-of-fit for specific experiments and the PDFs,
and in particular the correlations between some experimental data sets
and the Higgs boson cross section,
will be discussed further in Sections 3 and 4.

\begin{table}[p]
\begin{tabular}{|l|l|l|l||l|l|}
\hline
\textbf{ID\# } & \textbf{Experimental data set} & $N_{pt}$ & $\chi_{e}^{2}/N_{pt}$ & prob. & $S$ \tabularnewline
\hline
\hline
101  & BCDMS $F_{2}^{p}$ \cite{Benvenuti:1989rh} & 339 & 1.16 & 0.97 & 1.95 \tabularnewline
\hline
102  & BCDMS $F_{2}^{d}$ \cite{Benvenuti:1989fm} & 251 & 1.18 & 0.98 & 1.97 \tabularnewline
\hline
103  & NMC $F_{2}^{p}$ \cite{Arneodo:1996qe}     & 201 & 1.68 & 1.00 & 5.72 \tabularnewline
\hline
104  & NMC $F_{2}^{d}/F_{2}^{p}$ \cite{Arneodo:1996qe} & 123 & 1.20 & 0.93 & 1.51 \tabularnewline
\hline
108  & CDHSW $F_{2}^{p}$ \cite{Berge:1989hr}     & 85  & 0.82 & 0.12 & -1.18 \tabularnewline
\hline
109  & CDHSW $F_{3}^{p}$ \cite{Berge:1989hr}     & 96  & 0.79 & 0.06 & -1.53 \tabularnewline
\hline
110  & CCFR $F_{2}^{p}$ \cite{Yang:2000ju}       & 69  & 0.97 & 0.46 & -0.10 \tabularnewline
\hline
111  & CCFR $xF_{3}^{p}$ \cite{Seligman:1997mc}  & 86  & 0.40 & 0.00 & -5.19 \tabularnewline
\hline
124  & NuTeV neutrino dimuon SIDIS \cite{Mason:2006qa}& 38 & 0.78 & 0.16 & -0.99 \tabularnewline
\hline
125  & NuTeV antineutrino dimuon SIDIS \cite{Mason:2006qa} & 33 & 0.86 & 0.31 & -0.50 \tabularnewline
\hline
126  & CCFR neutrino dimuon SIDIS \cite{Goncharov:2001qe}      & 40 & 1.19 & 0.81 & 0.88 \tabularnewline
\hline
127  & CCFR antineutrino dimuon SIDIS \cite{Goncharov:2001qe}  & 38 & 0.69 & 0.07 & -1.46 \tabularnewline
\hline
140  & H1 $F_{2}^{c}$ \cite{Adloff:2001zj}      &  8 & 1.13 & 0.66 & 0.42 \tabularnewline
\hline
143  & H1 $\sigma_{r}^{c}$ for $c\bar{c}$ \cite{Aktas:2004az,Aktas:2005iw}  & 10 & 1.60 & 0.90 & 1.28 \tabularnewline
\hline
145  & H1 $\sigma_{r}^{b}$ for $b\bar{b}$  \cite{Aktas:2004az}      & 10 & 0.70 & 0.28 & -0.60 \tabularnewline
\hline
146  & H1 $F_{2}^{c}$ from $D^{*}$ \cite{Aktas:2005iw}       & 25 & 0.94 & 0.45 & -0.12 \tabularnewline
\hline
156  & ZEUS $F_{2}^{c}$ \cite{Breitweg:1999ad}  & 18 & 0.72 & 0.20 & -0.83 \tabularnewline
\hline
157  & ZEUS $F_{2}^{c}$ \cite{Chekanov:2003rb}  & 27 & 0.59 & 0.05 & -1.68 \tabularnewline
\hline
159  & Combined HERA1 NC and CC DIS \cite{Aaron:2009wt}  & 579 & 1.06 & 0.85 & 1.05 \tabularnewline
\hline
169  &  H1 $F_L$  \cite{Collaboration:2010ry}            & 9 & 1.71 & 0.92 & 1.40 \tabularnewline
\hline
201  & E605 Drell-Yan process, $\sigma(pA)$ \cite{Towell:2001nh} & 119 & 0.80 & 0.05 & -1.62 \tabularnewline
\hline
203  & E866 Drell Yan process, $\sigma(pd)/(2\sigma(pp))$ \cite{Webb:2003ps} & 15 & 0.60 & 0.12 & -1.16 \tabularnewline
\hline
204  & E866 Drell-Yan process, $\sigma(pp)$ \cite{Webb:2003ps}  & 184 & 1.27 & 0.99 & 2.44 \tabularnewline
\hline
225  & CDF Run-1 $W$ charge asymmetry \cite{Abe:1996us}           & 11  & 1.19 & 0.71 & 0.55 \tabularnewline
\hline
227  & CDF Run-2 $W$ charge asymmetry \cite{Acosta:2005ud}           & 11 & 1.02 & 0.58 & 0.19 \tabularnewline
\hline
231  & D0 Run-2 $W e\nu_{e}$ charge asymmetry \cite{Abazov:2008qv} & 12 & 2.09 & 0.99  & 2.18 \tabularnewline
\hline
234  & D0 Run-2 $W \mu\nu_{\mu}$ charge asymmetry \cite{Abazov:2007pm} & 9 & 1.20 & 0.71  & 0.55 \tabularnewline
\hline
260  & D0 Run-2 $Z$ rapidity distribution \cite{Abazov:2006gs}       & 28 & 0.58 & 0.04  & -1.77 \tabularnewline
\hline
261  & CDF Run-2 $Z$ rapidity distribution \cite{Aaltonen:2010zza}   & 29 & 1.60 & 0.98  & 2.03 \tabularnewline
\hline
268  & ATLAS W and $Z$ production    \cite{Georges:2012aa}               & 41 & 0.87 & 0.29  & -0.56 \tabularnewline
\hline
504  & CDF Run-2 inclusive jet production \cite{Aaltonen:2008eq}   & 72 & 1.39 & 0.98  & 2.12 \tabularnewline
\hline
514  & D0 Run-2 inclusive jet production \cite{Abazov:2008hua}     & 110 & 1.03 & 0.59  & 0.24 \tabularnewline
\hline
535  & ATLAS single inclusive jet data with R=0.6 \cite{Georges:2012ab}   & 90 & 0.70 & 0.01  & -2.21 \tabularnewline
\hline
\end{tabular}
\caption{Experimental data sets employed in our analysis.
$N_{pt}$ = the number of data points;
$\chi_{e}^{2}/N_{pt}$ = the value for the global minimum.
The total number of data points is 2797.
The fifth column is the cumulative probability that a true
chi-square distribution
with $N_{pt}$ points would have $\chi^{2} \leq \chi_{e}^{2}$.
The final column is the equivalent Gaussian variable $S$,
which is discussed in Sec. IV.
\label{tab:EXP_bin_ID}
}
\end{table}

Beyond the determination of the ``best fit'', the next goal of the global analysis is to determine the uncertainties on the PDFs.
Two methods for PDF uncertainty estimation have been used by the CTEQ Collaboration: the Hessian method and the Lagrange multiplier method.
The Hessian method~\cite{Pumplin:2000vx,Pumplin:2001ct}
is based on standard error propagation;
it relies to some extent on a quadratic Gaussian approximation.
The Lagrange multiplier method~\cite{Stump:2001gu} does not depend on the quadratic approximation,
and therefore is more robust~\cite{Pumplin:2002vw}.

The Hessian method does not focus on any particular prediction.  The PDF uncertainty for any observable can
be calculated in this method using ``error PDFs''.
It relies on the assumption that the behavior of the global $\chi^{2}$ function
is quadratic within the range of the uncertainties for all the PDF fitting parameters.
This assumption cannot be valid for large variations of the PDFs,
so if the uncertainty of a prediction is large, we may question the validity of the Hessian method.
On the other hand, the LM method focuses on a particular observable, and finds the limit to goodness of fit
as that observable moves away from its central value.
The LM method can be used to test whether the Hessian method is valid for that observable.

We review these two methods in succession below.

\subsection{Hessian Method}\label{HM}

The Hessian method~\cite{Pumplin:2001ct} for the analysis of PDF uncertainty
starts with the Hessian matrix
\begin{equation}
H_{ij}=\frac{1}{2} \left( \frac{\partial^{2}{\chi}^{2}}
 {\partial{a}_{i}\partial{a}_{j}} \right) _{0}
\end{equation}
evaluated at the minimum of $\chi^2$.
$H_{ij}$ determines the behavior of $\chi^{2}(a)$ near the central fit, with the PDF parameters $\{a_{0}\}$.
We next calculate the eigenvectors of $H_{ij}$.
For each eigenvector we compute two displacements from $\{a_{0}\}$
(in the $+$ and $-$ directions along the vector)
denoted by $\{a^{+}_{k}\}$ and $\{a^{-}_{k}\}$ for the $k$-th eigenvector.
The distance from $\{a_{0}\}$ is defined by
${\chi}^{2}_{\pm} = {\chi}_{0}^{2} + T^{2}$,
where $T$ specifies the \emph{tolerance}.
The appropriate choice of tolerance $T$ cannot be decided without further,
more detailed, analyses of the quality of the global fits.

The choice of tolerance $T$  has been debated for a long time.
After studying a number of examples~\cite{Stump:2001gu, Pumplin:2001ct, Pumplin:2002vw},
the CTEQ group has concluded that a rather large tolerance, $T \sim 10$,
represents a realistic estimate of the full PDF uncertainty at the $90\%$ confidence level (C.L.).
Other PDF research groups have made different choices, or have used other quantities to measure the goodness of fit.
In addition, for the CTEQ-TEA global analyses, we do not accept that the naive condition $T<10$ is sufficient in itself for setting the uncertainties of the PDFs and their predictions.
We also need to test whether any individual data sets would disagree with the candidate PDFs.
For that purpose, we add ``tier-2 penalty'' terms to the global $\chi^{2}$ function
and demand that the {\em combination} not be too large\footnote{The tier-2 penalty is described in more detail below.}.
We have found this procedure to give a satisfactory estimate of the agreement between data and theory, and use it
as the basis for our uncertainty analyses.  Whether the tier-2 penalties will have a significant impact
must be checked for each application.

Comparisons of PDF uncertainties from different PDF groups, such as the comparisons in the benchmark
calculations in Ref.\ \cite{benchmark} show that the choice $T=10$ with tier-2 penalties gives results
that are generally comparable to PDF uncertainties calculated by other methods.
In any case, our purpose here is to compare Hessian and LM results.
Since we apply the same $T=10$ criterion in both cases, we shall see if the Hessian method is trustworthy.

We view $T \sim 10$ as a measure to estimate the possibly large (but not unreasonable) error coming from the many sources of uncertainties in global analysis,
in the nature of a $90\%$ C.L., rather than Gaussian standard deviation.
Very often, the comparison of the PDF uncertainty to the experimental data is performed at a $68\%$ C.L., which should be converted from the result obtained at the $90\%$ C.L. by a scaling
factor of $1/1.645$ when using the CT PDF sets.
We note that the total $90\%$ C.L. uncertainty on typical observables has contributions from several sources, including the experimental, theoretical,
PDF parametrization, and procedural uncertainties.\footnote{For simple observables,
such as the $\overline{\rm MS}$ charm quark mass determined from the global fit, the PDF uncertainty based on the $T\sim 10$ criterion can be demonstrated to be close to
the $90\%$ C.L. uncertainty from a combination of sources~\cite{Gao:2013wwa}.}

One can show that for ideal Gaussian errors,
the {\em symmetric} uncertainty $\delta X$ for any quantity $X$ that depends on PDFs
can be expressed as
\begin{equation}
\left(\delta{X}\right)^{2} = T^{2} \sum_{i,j} \left(H^{-1}\right)_{ij} \frac{%
\partial{X}}{\partial{a}_{i}} \frac{\partial{X}}{\partial{a}_{j}};
\end{equation}
or, in terms of the eigenvector basis sets,
\begin{equation}  \label{eq:ME1}
\left(\delta{X}\right)^{2} = \frac{1}{4}\sum_{k=1}^{n} \left[%
X(a^{+}_{k})-X(a^{-}_{k})\right]^{2},
\end{equation}
which is called the master equation in Ref.~\cite{Pumplin:2001ct}.
However, Eq. (\ref{eq:ME1}) is based on the following approximations:
${\chi}^{2}(a)$ is assumed to be a quadratic function of the parameters $\{a\}$, and $X(a)$ is assumed to be a linear function of $\{a\}$ around the central fit.
These approximations are not strictly valid in general.
Therefore, to better take any nonlinearities into account, we calculate \emph{asymmetric errors}
from the eigenvector basis sets~\cite{Nadolsky:2001yg,Lai:2010vv}.\footnote{%
A more complete explanation of the CT global analysis
was given in Ref.\ \cite{ct10nnlo}.}

\subsection{Lagrange multiplier Method}\label{LM}

The Lagrange multiplier (LM) method~\cite{Stump:2001gu} is complementary to the Hessian method.
The idea of this method is to make \emph{constrained fits}.
The trick is to introduce a Lagrange multiplier variable $\lambda$,
and to minimize the function
\begin{equation}\label{eq:lmf}
F(a,\lambda)= \chi^2(a) + \lambda \left[X(a) - X(a_0)\right]
\end{equation}
at fixed values of $\lambda$.
Again, $X(a)$ is the observable that we are trying to predict,
and $X(a_{0})$ is the central prediction.
For a given value of $\lambda$, the values of the parameters, $\{\overline{a}\}$,
at the minimum correspond to the best fit with the corresponding value of the observable constrained to $X = X(\overline{a})$.
That is,  $\chi^{2}(\overline{a})$ is the minimum value of $\chi^{2}$ with the constraint $X = X(\overline{a})$.

The result of the LM calculation is a series of constrained fits.
We do the calculations for many values of $\lambda$,
and thereby trace out the constrained fit as a function of $X$.
A graph of $\chi^{2}(\overline{a})$ versus $X(\overline{a})$ shows the variation of $\chi^2$ around the minimum $\chi^2_0$ for possible alternative fits that give different values of the observable $X$.
The uncertainty of the prediction of $X$ is then obtained from the condition $\chi^{2}(\overline{a})-\chi_{0}^{2} = T^{2}$,
where $T$ is again the \emph{tolerance}.
We thus obtain the $90\%$ C.L. uncertainties in the observable, specified by the maximum value
$X_{\rm max} = X(a_0) + \delta{X}_{+}$
and the minimum value
$X_{\rm min} = X(a_0) - \delta{X}_{-}$.
To compare with the result obtained from the Hessian method at the $68\%$ C.L.,
we estimate the uncertainty derived from the LM method by choosing
$\chi^2-\chi_o^2=(T/1.645)^2=T^2/2.71\,$.
This prescription may break down when the quadratic approximation is badly violated.

In general, the LM method for calculating $\delta X_\pm$ is less convenient than the Hessian method.
In the Hessian method, once the eigenvector PDF sets are computed, the uncertainty for any
observable can be straightforwardly calculated without redoing the global analysis.
In the LM method, the global analysis minimization procedure has to be run separately for every observable of interest.
On the other hand, in contrast to the Hessian method, the LM method does not assume $\chi^{2}(a)$ and $X(a)$ to have quadratic and linear dependence on $\{a\}$, respectively, around the minimum.
Moreover, the LM method provides more information about $X$.
It determines the full functional dependence of $\chi^2(\overline a)$ on $X(\overline a)$,
and the confidence intervals on $X$ can be recomputed from this dependence if a tolerance other than $T=10$ is prescribed.  (The standard error PDFs in the Hessian method
depend on the choice of $T$.)

A comparison of the LM and Hessian methods will indicate the degree to which these approximations are reasonable, and to which the Hessian method accurately predicts the uncertainties of the Higgs boson cross section,
at both the $90\%$ C.L. and $68\%$ C.L..

We emphasize that interest in the Lagrange multiplier method should not be limited to the CT10 global analysis.
Its general impact is that it provides a check on the validity of the commonly used Hessian method.
Furthermore, it can provide detailed information on the correlations between the values of different observables, in the manner shown in Sections. III and IV.
In the context of the Higgs cross section, we shall see how correlations between $\alpha_{s}$ and the PDFs affect uncertainties
on the prediction of $\sigma_{H}$ in Sec.\ III.

\subsection{Special treatment of the uncertainty with respect to $\alpha_{\rm s}$}

In addition to the PDF uncertainties, the uncertainty in the value of the QCD coupling $\alpha_{\rm s}(M_{Z})$ will also contribute to the uncertainty of the prediction of the observable $X$
(and to errors on the theory values $\{T_{\nu}(a)\}$).
Various approaches have been developed to deal with this particular source of uncertainty, which were reviewed in our previous study on this subject~\cite{Lai:2010nw}.
In this work we expand on that study.
Following the PDF4LHC~\cite{PDF4LHC} working group guidelines, we take
\begin{equation}
\alpha_{s}(M_{Z}) = \alpha_{\rm WA} \pm \delta\alpha_{\rm WA}\label{eq:alphas}
\end{equation}
where the current ``world-average'' central value is $\alpha_{\rm WA} = 0.118$
with a $90\%$ C.L. of $\delta\alpha_{\rm WA} = 0.002$.
This corresponds to a $68\%$ C.L. uncertainty of $\pm0.0012$.
The world-average value~\cite{Bethke,PDG} is obtained
from various experimental and theoretical results,
including the deep-inelastic scattering (DIS) data that are also used in the current global analysis.
The errors in $\alpha_s(M_Z)$ given in Eq.~(\ref{eq:alphas}), advocated by the PDF4LHC working group and used in the present work,
are larger than the world-average errors given by the Particle Data Group (PDG)~\cite{PDG}; however, they are still substantially smaller than the errors on $\alpha_s(M_Z)$ obtained from the global analysis alone.

The standard method for including uncertainties from $\alpha_s(M_Z)$ in the Hessian method was outlined in Ref.~\cite{Lai:2010nw}.
In this method, two additional PDF sets
(obtained from the best fits with values of $\alpha_s(M_Z)=
\alpha_{\rm WA} \pm \delta\alpha_{\rm WA}$) are used to calculate the uncertainty on
the observable $X$ due to the uncertainty of $\alpha_s(M_Z)$.
This is then added in quadrature to the uncertainty on $X$ due to uncertainties of the PDFs.

In the LM method the additional uncertainty from $\alpha_s(M_Z)$ can be obtained by treating
$\alpha_{s}(M_{Z})$ as another fitting parameter.
Using the current CT10H NNLO global analysis alone, we obtain a determination of $\alpha_s(M_Z)$ that is consistent with the world-average result,
but smaller and with a much larger uncertainty of $\pm0.006$ at the $90\%$ C.L..
To include the additional (and stronger) constraints on $\alpha_s(M_Z)$ from the world-average analysis in the LM method,
we add a penalty to the global $\chi^{2}$ function,
\begin{equation}\label{eq:kappa}
\chi^{2} \rightarrow
\chi^{2} + \kappa ~ \left[
(\alpha_{s}(M_{Z})-\bar\alpha)/\delta\bar\alpha
\right]^{2},
\end{equation}
where $\bar\alpha=0.1186$ and $\delta\bar\alpha=0.0021$.
They can be interpreted as the PDF4LHC world-average value and errors on $\alpha_s(M_Z)$, but with the DIS data removed, so as not to double-count in the global analysis.
The $\kappa$-penalty term incorporates the additional world-average constraints on $\alpha_s(M_Z)$ in a manner analogous to an additional data set in the global $\chi^2$ function.
We note that the value of $\bar\alpha=0.1186$ is consistent with the quoted value of $\alpha_s$, obtained by leaving out the DIS data from the world-average analysis in Ref.~\cite{Bethke}.
Finally, the weight factor $\kappa$ is chosen to be equal to the tolerance-squared,
$\kappa=T^2=100$, consistent with the interpretation of $\delta\bar\alpha$ as the $90\%$ C.L. uncertainty in $\alpha_s(M_Z)$ arising from the  world-average constraints, after excluding  the DIS data.

In Ref.~\cite{Lai:2010nw} it was shown that the Hessian method and the LM method of including the $\alpha_s(M_Z)$ uncertainties agree in the quadratic approximation.
In Appendix 1 we re-derive this for the special case of a single observable $X$
(more specifically, for the Higgs boson cross section).
In particular, we take into account more carefully the various origins of the constraints on $\alpha_s(M_Z)$, and show that the two methods agree in the quadratic approximation,
if the world-average constraints on $\alpha_s(M_Z)$ are included in the LM method as described in the last paragraph.
Thus, a comparison of the combined PDF$+\alpha_s$ error determined from the Hessian method to that determined from the LM method can provide a further check on the Hessian prescription of adding PDF and $\alpha_s$ errors in quadrature when the quadratic approximation is valid.

\subsection{tier-2 Penalties}
In both the Hessian and LM approaches,
it is often found that the global $\chi^2$ by itself, summed over all experiments,
is not an adequate indicator of goodness-of-fit.
It is possible to have a low value of the global $\chi^{2}$,
for which one or a few experiments have poor fits to the theory,
but balanced by other experiments with unexpectedly good agreement with the theory.
In the large data sample used in the CTEQ analysis,
we expect to encounter unacceptable fits of this kind,
so it is important to check the agreement with each individual experiment.
Ideally, no single experiment should conflict so strongly with the set of test PDFs that it would rule out that test set on its own,
even if the \emph{global} $\chi^{2}$ is acceptable.
Our  method for judging the agreement between theory and data is to add an extra ``tier-2 penalty'' to the global $\chi^2$ for every experiment that is fitted.
The tier-2 penalty is a contribution for each experiment which is given by a function that increases rapidly if that experiment's data deviate from the theory predictions beyond the $90\%$ C.L..
The tier-2 penalty was introduced in \cite{Lai:2010vv},
and its current technical implementation is extensively reviewed in Ref.~\cite{Dulat:2013hea}.

Thus, it is the $\chi^{2} + \mbox{tier-2 penalties}$, rather than the global $\chi^{2}$,
that serves as a figure of merit for the global fit.
When generating the PDF eigenvector sets,
we find that about a half of them are constrained by the growth of some tier-2 penalty ({\it i.e.},
one specific experiment),
and not by the growth of the global $\chi^2$.
Both the CT10 and CT10H PDF error sets have been generated with a tier-2 penalty.
When interpreting the PDF error on physics observables
such as the $gg\rightarrow H$ cross section,
we may ask whether the tier-2 penalty plays a significant role.
In Section IV, we shall examine this question in the context of the LM method in order to gain insights on the source of the PDF uncertainty on the Higgs boson cross section.

\section{Uncertainty of Higgs boson Cross Section from PDFs and $\alpha_s(M_Z)$}
\label{sec:Uncert}

In the current study, we are concerned primarily with the error estimate
on the Higgs boson production cross section $\sigma_H$ via the gluon fusion process,
induced by both the PDF and $\alpha_s$ uncertainties.
For the calculation of $\sigma_H$, we have utilized the NNLO code
HNNLO1.3~ \cite{hnnlo1,hnnlo2}
for a Higgs boson mass $M_H = 125\ {\rm GeV}$, with both the renormalization and
factorization scales fixed at $\mu=M_H$.\footnote{
At fixed order, the LHC Higgs Cross Section Working Group Collaboration~\cite{Dittmaier:2011ti}
has chosen a scale at $M_H/2$, leading to larger cross sections; cf.~Appendix 2.
Our conclusions about percentage uncertainties can be applied to either prediction.
}
In the following, we calculate and compare the uncertainties obtained in the
two different methods:  the Hessian method and the Lagrange multiplier method.

\subsection{The Hessian Calculation}
\label{sec:HessianUncert}

In the Hessian method,
the PDF uncertainties on the predictions of an observable
are calculated using a standard set of \emph{error PDFs}.
The uncertainty on the observable due to the $\alpha_s$ uncertainty is calculated
using the PDF $\alpha_s$ series.
The PDF and $\alpha_{s}$ errors are then combined in quadrature
to obtain the total uncertainty~\cite{Lai:2010nw}.
In Table \ref{tbl:xsecs} we list the predicted central values,
PDF uncertainties and PDF+$\alpha_s$ uncertainties,
at $90\%$ C.L. and $68\%$ C.L.,
for the cross sections of Higgs boson production
via gluon fusion at the LHC.
The error PDFs and $\alpha_{s}$ series of the CT10H NNLO global analysis
were used for the calculations, including the contribution of the tier-2 penalty to $\chi^2$.
Generally, the predictions from CT10H agree well with those from CT10;
cf.~Appendix 2.
To obtain the $68\%$ C.L. errors,
we used the standard prescription of scaling the $90\%$ errors
by a factor of $1/1.645$.

\begin{table}[h]
\begin{center}
\begin{tabular}{c|ccc}
\hline \hline
LHC & 7 TeV & 8 TeV & 14 TeV \\
\hline
$gg \to H$ (pb) with $90\%$ C.L. errors&
 $13.4^{+4.7(3.0)\%}_{-4.6(3.0)\%}$ &
 $17.0^{+4.8(3.2)\%}_{-4.7(3.1)\%}$ &
 $44.5^{+5.4(4.3)\%}_{-5.0(3.6)\%}$ \\
\hline
\ \ \ \ \ \ \ \ \ \ \ \ \ \ \ \ \ \ with 68\% C.L. errors&
 $13.4^{+2.9(1.8)\%}_{-2.8(1.8)\%}$ &
 $17.0^{+2.9(1.9)\%}_{-2.8(1.9)\%}$ &
 $44.5^{+3.3(2.6)\%}_{-3.0(2.2)\%}$ \\
\hline
\hline
\end{tabular}
\end{center}
\vspace{-2ex}
\caption{\label{tbl:xsecs}
Higgs boson production cross sections (in pb unit)
via gluon fusion channel at the LHC, with 7, 8 and 14 TeV center-of-mass energy.
The combined PDF and $\alpha_s$ uncertainties
and the PDF-only uncertainties (inside the parentheses),
at the $90\%$ C.L. and $68\%$ C.L.,
were calculated by the Hessian method with the
CT10H NNLO error PDF sets.
The uncertainties are expressed as the percentage of the central value,
and the PDF-only uncertainties are for $\alpha_s(M_Z)=0.118$.
}
\end{table}

\subsection{The LM calculation of the PDF uncertainty of $\sigma_{H}$}
\label{sec:LMUncert}

We first perform the calculations using the LM with fixed
$\alpha_{\rm s}(M_{Z}) = \alpha_{\rm WA} = 0.118$, in order to
obtain the uncertainty in the prediction of the Higgs boson cross section purely due to the
PDF uncertainties.  Again, this analysis is for Higgs boson production
through gluon fusion
for $pp$ collisions at energies $\sqrt{s} = $ 7, 8 and 14 TeV.

\begin{figure}[H]
\begin{center}
\includegraphics[width=0.47\textwidth]{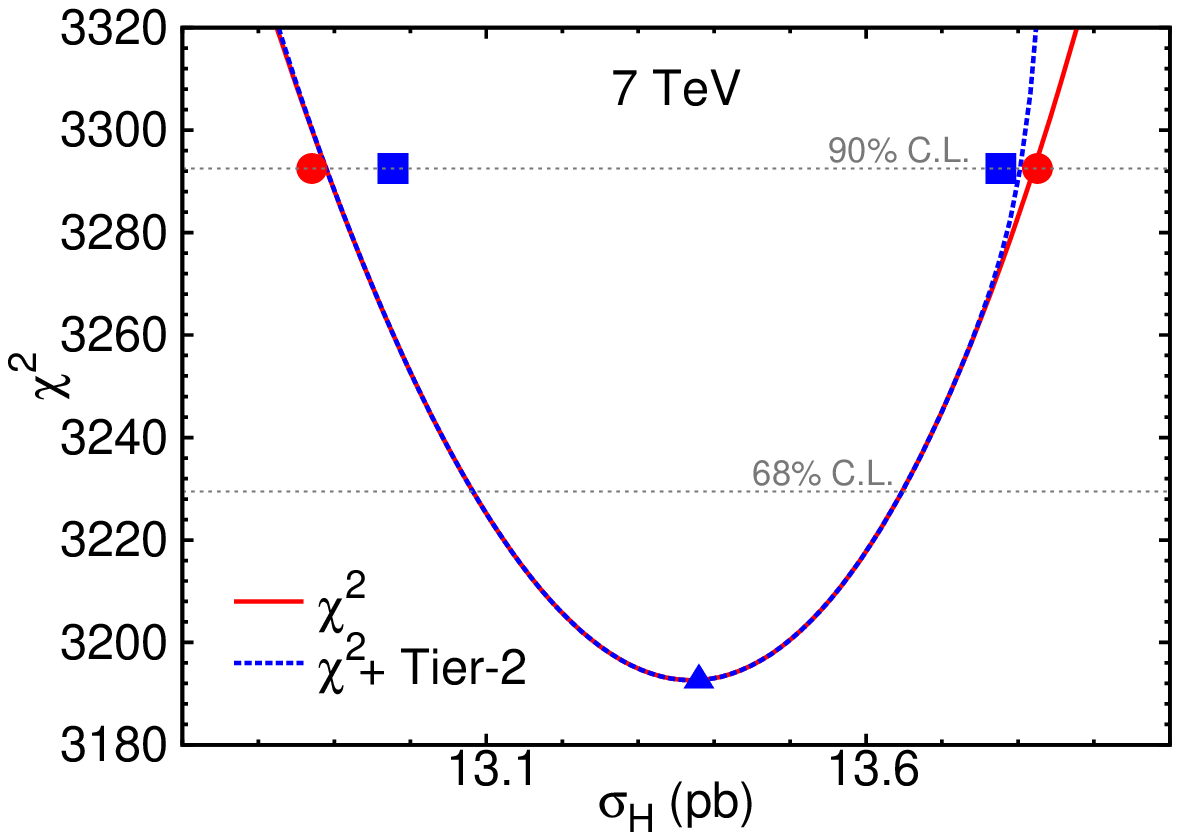}
\includegraphics[width=0.47\textwidth]{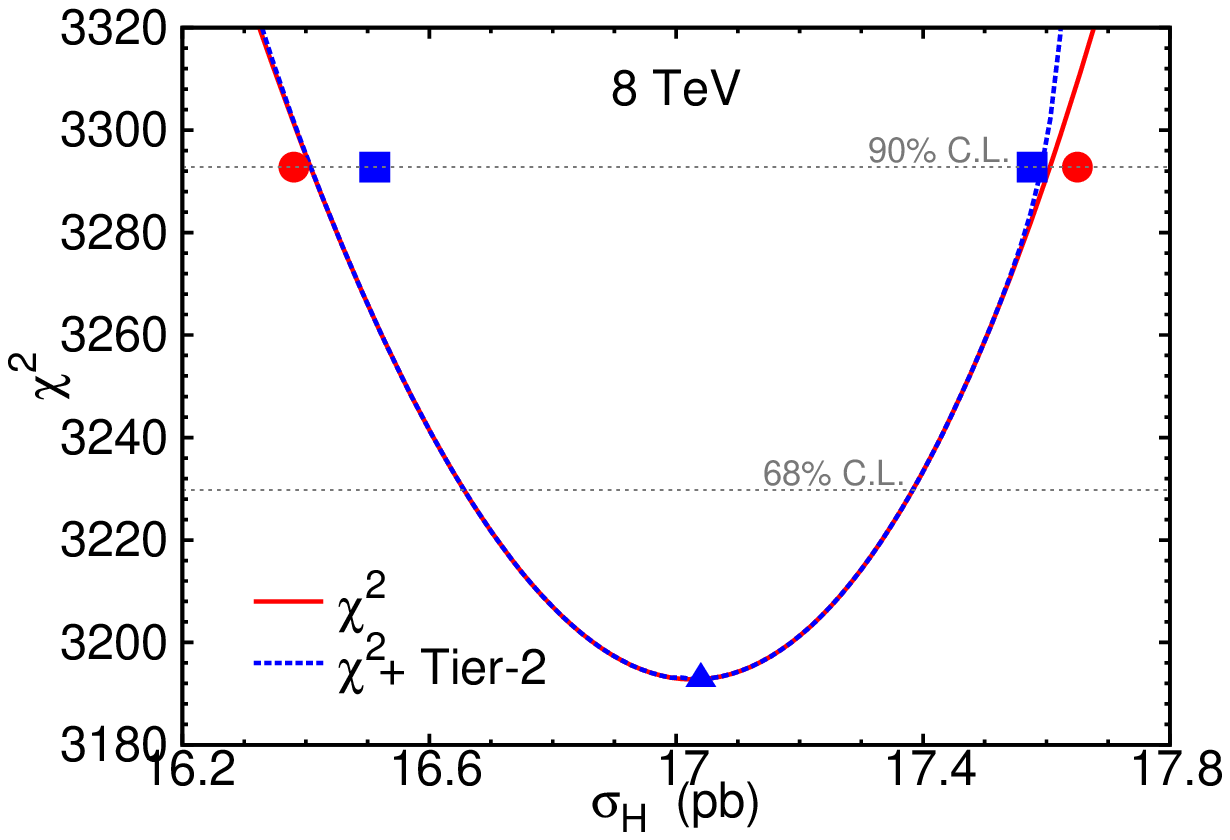}
\includegraphics[width=0.47\textwidth]{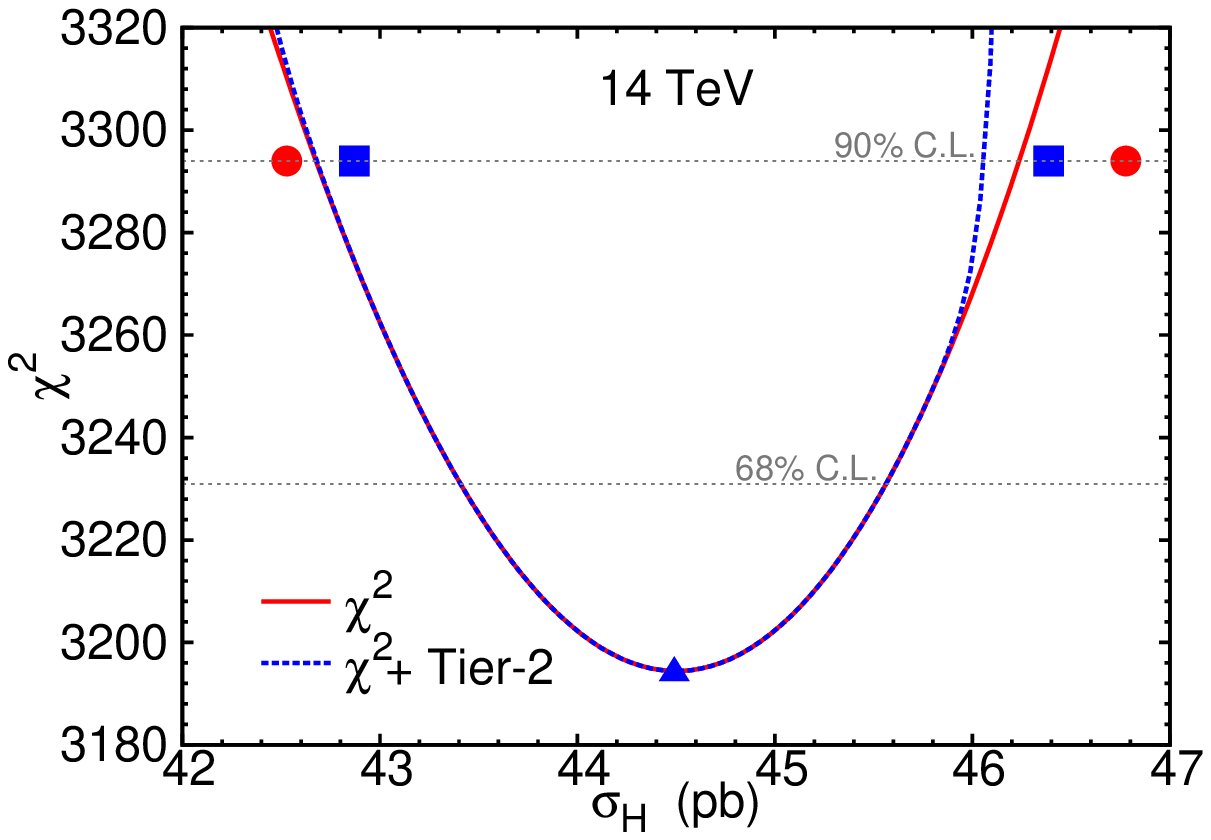}
\end{center}
\caption{$\chi^{2}$ versus $\sigma_{H}$ with $\alpha_{s}(M_{Z}) = 0.118$.
The constrained minimum of $\chi^{2}$ is plotted as a function
of the predicted cross section $\sigma_{H}$
for Higgs boson production via gluon fusion channel at the LHC, for
$\sqrt{s} = 7, 8~ {\rm and}~ 14\ {\rm  TeV}$.
The constrained fits without and with the tier-2 penalties are shown
as red solid and blue dashed curves, respectively.
The red circles and blue boxes indicate the $90\%$ C.L.  errors
obtained from the Hessian
method without and with the tier-2 penalties, respectively.
\label{fig:LMparab}}
\end{figure}

Figure \ref{fig:LMparab} shows the results of the constrained fits,
for the three center-of-mass energies, represented
by the curves of $\chi^{2}(\overline{a})$
versus $\sigma_{H}(\overline{a})$.
The minimum is at the central prediction $\sigma_{H}(a_0)$.
The asymmetric errors $(\delta\sigma_{H})_{\pm}$ at the $90\%$ C.L. are determined
from this curve by the tolerance $T$,
\begin{eqnarray}
\chi^{2}(\overline{a}) &=& \chi^{2}(a_{0}) + T^{2} ~,\\
\sigma_{H}(\overline{a}) &=& \sigma_{H}(a_{0}) \pm (\delta\sigma_{H})_{\pm} ~.
\end{eqnarray}
Similarly, the 68\% C.L. is obtained from
$\Delta\chi^2=\chi^2(\bar{a})-\chi^2(a_0)=(T/1.645)^2$.  Using $T=10$, we have indicated
the $90\%$ C.L. and $68\%$ C.L. by the upper and lower horizontal lines, respectively, in each of the
plots.

The red solid curves, which are approximately parabolic, show $\chi^2$ versus $\sigma_{H}$;
the blue dashed curves are for $\chi^2 + $tier-2 penalty versus $\sigma_H$.
The two curves are almost identical over much of the range plotted.  They only begin to diverge
at large values of $|\sigma_{H} - \sigma_{H}(a_0)|$, where one or more experimental
data sets can no longer be satisfactorily fit, resulting in a large tier-2 penalty.
The blue triangles on the curves in Fig.~\ref{fig:LMparab} are the central values of $\sigma_H$, and
the blue boxes are the
upper and lower $90\%$ C.L. limits, calculated in the Hessian method and listed in
Table~\ref{tbl:xsecs}.  The red circles correspond to the same quantities, also calculated in the Hessian method,
but without including the tier-2 penalties.  They are plotted at the vertical value of $\Delta\chi^2=
T^2=100$, and are to be compared to the curves calculated by the LM method.

By comparing the red solid curves and the red circles,
and noting the parabolic nature of the red solid curves,
we note that the quadratic approximation works well for $\sigma_H(gg \to H)$,
up to the tolerance bounds of $T^2=100$.
Also, the LM and Hessian methods, without including the tier-2 penalty, give comparable
estimates of the PDF errors on $\sigma_H$.
(The two error estimates without including the tier-2 penalty do differ somewhat more at 14 TeV.)
After including the tier-2 penalties in both the LM and Hessian methods, shown by the blue dashed curves
and blue boxes, we find that the error predictions become more asymmetric.
However, the differences in the error estimates
by the four methods shown in Fig.~\ref{fig:LMparab}
are still considerably smaller than the error estimates themselves.
Table \ref{tbl:LMtable}
gives the numerical values of the central predictions and asymmetric errors for $\sigma_{H}$, obtained from the LM
method, with the tier-2 penalties included.
Comparing Tables~\ref{tbl:xsecs} and~\ref{tbl:LMtable}, we find that
the PDF uncertainty estimates predicted by both methods
are similar.

A technical detail in both our LM and Hessian calculations
concerns the normalizations of the fixed target DIS experiments
BCDMS, CDHSW and CCFR.
These experiments have large numbers of data points,
and rather small quoted normalization uncertainties
($3\,$\%, $1\,$\%, and $2.6\,$\% respectively).
We found in previous CTEQ global analyses,
e.g., in calculating the uncertainty of $W^{\pm}$ or $Z^{0}$
production cross sections at the Tevatron,
that allowing these normalizations to vary beyond
their published standard deviations could produce
fits with fairly small $\chi^{2}$,
for quite large deviations of $\sigma_{W}$ or $\sigma_{Z}$.
But these fits were not acceptable because they required large shifts
in the normalizations of the DIS data~\cite{Stump:2001gu}.
For this reason we have chosen both in the past and in the current Higgs study
(including the published CT10 Hessian set)
to fix the normalizations of these three experiments at their best fit values.

\begin{table}
\begin{center}
\begin{tabular}{c|ccc}
\hline \hline
LHC & 7 TeV & 8 TeV & 14 TeV \\
\hline
$\sigma_H(gg \to H)$ (pb) with $90\%$ C.L. errors&
 $13.4^{+3.2\%}_{-3.7\%}$ &
 $17.0^{+3.2\%}_{-3.7\%}$ &
 $44.5^{+3.5\%}_{-4.1\%}$\\
\hline
\ \ \ \ \ \ \ \ \ \ \ \ \ \ \ \ \ \ with 68\% C.L. errors&
 $13.4^{+2.0\%}_{-2.2\%}$ &
 $17.0^{+2.0\%}_{-2.3\%}$ &
 $44.5^{+2.2\%}_{-2.4\%}$\\
\hline
\hline
\end{tabular}
\end{center}
\vspace{-2ex}
\caption{\label{tbl:LMtable}
Higgs boson production cross sections (in pb unit)
via gluon fusion channel  at the LHC, with 7, 8 and 14 TeV center-of-mass energy.
The PDF uncertainties
at the $90\%$ C.L. and $68\%$ C.L.
were calculated by the  Lagrange multiplier method in the CT10H analysis
with fixed $\alpha_s(M_Z)=0.118$.
The uncertainties are expressed as the percentage of the central value.
}
\end{table}

\subsection{The LM calculation of the combined PDF$+\alpha_s$ uncertainty of $\sigma_H$}

In the previous subsection, we presented results using the LM method, while treating
$\alpha_{s}(M_{Z})$ as a fixed parameter,
equal to the current world-average value.
We now consider the combined PDF and $\alpha_s(M_Z)$ effects
in the LM method, by including the world-average constraints
on $\alpha_s(M_Z)$ directly in the $\chi^2$ function, using Eq.~(\ref{eq:kappa})
and treating $\alpha_s(M_Z)$ as an additional fitting parameter.
In practice,
we select $\alpha_{s}(M_{Z})$ from a set of discrete values and repeat
a LM scan of $\chi^2$ for each selected $\alpha_{s}(M_{Z})$;
that determines the constrained $\chi^{2}(\overline{a})$ versus
$\sigma_{H}(\overline{a})$ in a range of $\alpha_{s}(M_Z)$.
(The term with $\kappa=100$ introduced in
Eq.\ (\ref{eq:kappa}) to specify the world-average constraints
on $\alpha_s(M_Z)$ is now included as a part of $\chi^{2}$.)
We perform the calculations for a series of values of
$\alpha_{s}(M_Z)$.
Then, we have $\chi^{2}$ as a function of $(\alpha_{s}(M_Z),\sigma_{H})$,
and we can trace out contours of $\chi^{2}$
in the $(\alpha_{s},\sigma_{H})$ plane.

\begin{figure}[H]
\begin{center}
\includegraphics[width=0.47\textwidth]{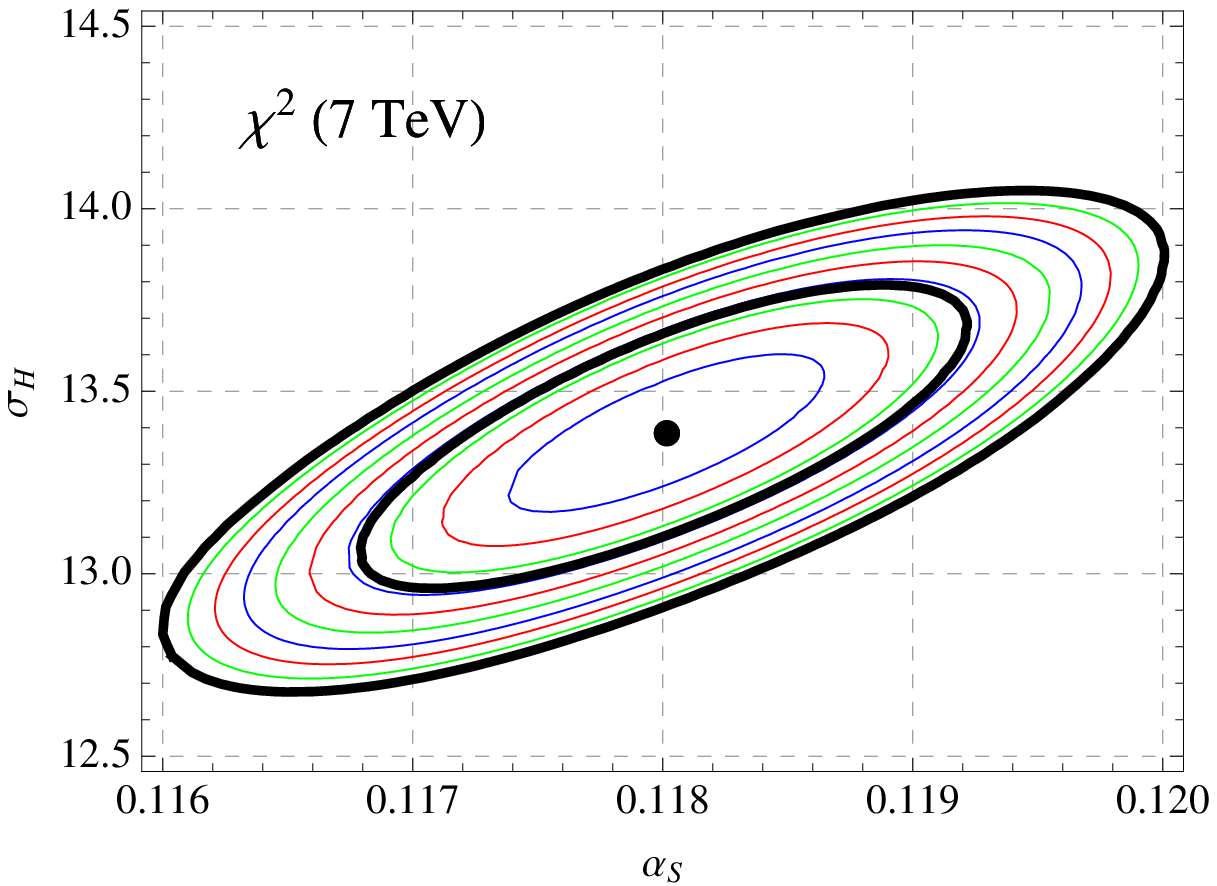} \,\,\,
\includegraphics[width=0.47\textwidth]{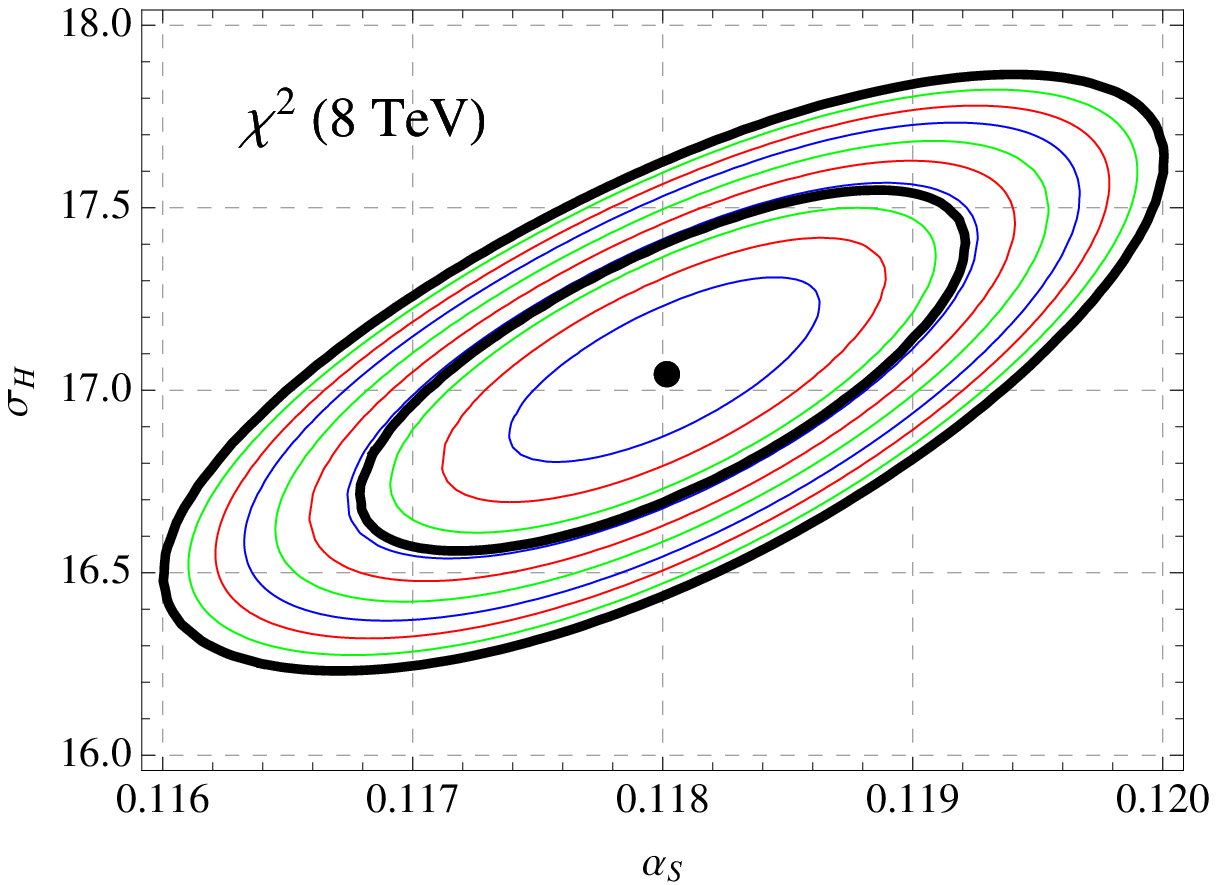}
\includegraphics[width=0.47\textwidth]{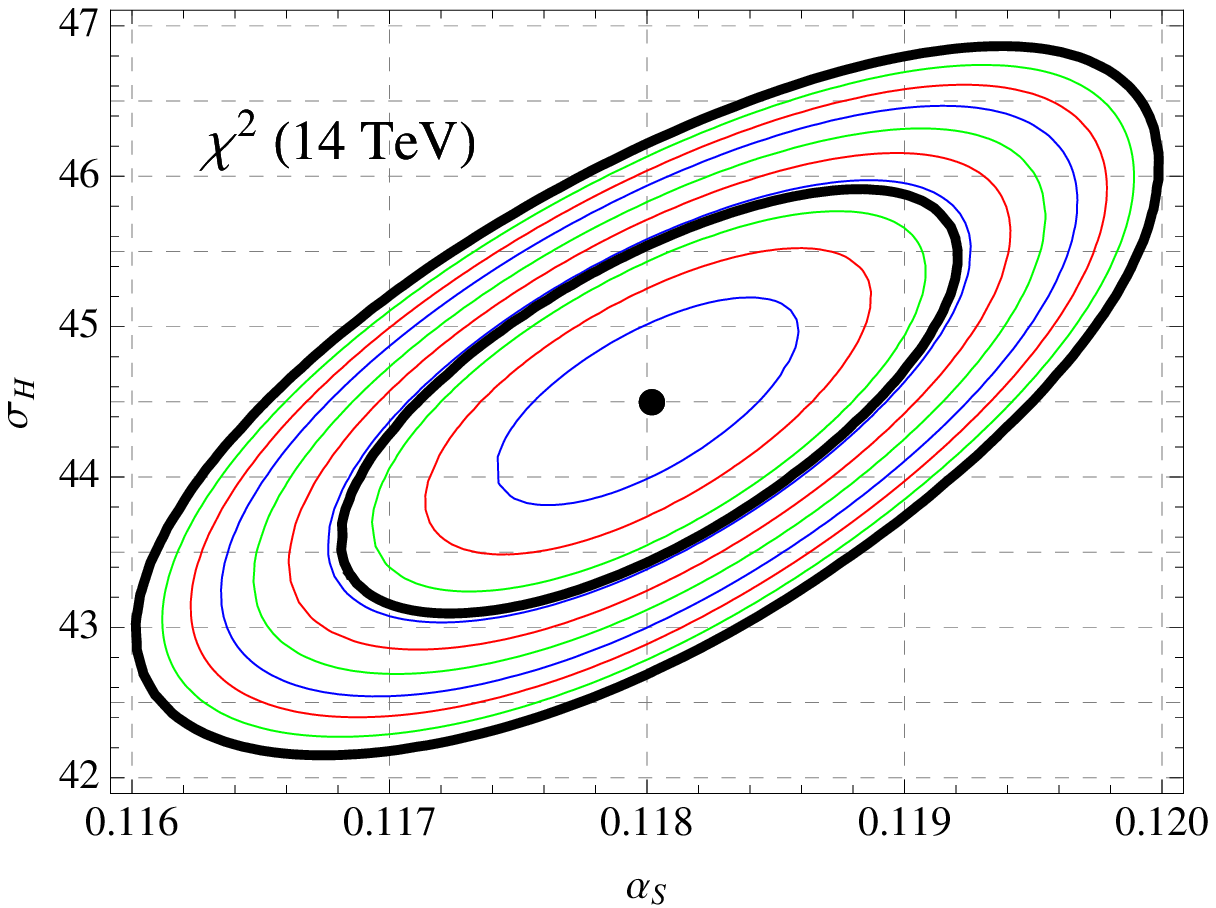}
\end{center}
\caption{Contour plots of $\chi^{2}(\overline{a})$
in the $(\alpha_{s},\sigma_{H})$ plane,
for $\sigma_H$ (in pb unit) at the LHC, with 7, 8 and 14 TeV.
The thick black outer and inner contours are at
$\Delta\chi^2=100$ and $100/(1.645)^2$, respectively, for the $90\%$ C.L.
and $68\%$ C.L..  The thin colored contours are at intervals in $\chi^2$ of 10.
\label{fig:contours}}
\end{figure}

Figure \ref{fig:contours} shows the
contour plots of $\chi^{2}(\overline{a})$
in the $(\alpha_{s},\sigma_{H})$ plane,
for $\sigma_H$ at the LHC with $\sqrt{s} = $ 7, 8 and 14 TeV.
A contour here is the locus of points in the $(\alpha_{s}, \sigma_{H})$ plane
along which the constrained minimum of $\chi^{2}$ is constant.
Note that we have not included the tier-2 penalty in the calculation of $\chi^2$ for
Fig.~\ref{fig:contours}.
We see from these figures that the values of
$\sigma_{H}$ and $\alpha_{s}(M_{Z})$ are strongly correlated,
as expected, given the strong dependence of the $gg$ fusion cross
section on $\alpha_s(M_Z)$ at NNLO.
Larger values of $\alpha_{s}(M_{Z})$ correspond to
larger values of $\sigma_{H}$ for the same goodness-of-fit
to the global data, even though there is a partially compensating decrease of the
$gg$ luminosity.

The contour with $\chi^{2}-\chi_{0}^{2} = T^{2}$
is particularly interesting, because it
represents our estimate for the \emph{correlated uncertainties}
of $\alpha_{s}$ and $\sigma_{H}$ at the $90\%$ C.L..
This contour gives $\alpha_s(M_Z)=0.118\pm0.002$,
as we expected from the discussion in Section II.  By finding the
extreme values of $\sigma_{H}$ along the contour,  we obtain the combined PDF~$+ \alpha_s$ errors on the Higgs cross section, which are displayed inTable \ref{tbl:contour100}.
Similar results at the $68\%$ C.L. are also shown in both
Fig.~\ref{fig:contours} and Table \ref{tbl:contour100}.

\begin{table}
\begin{center}
\begin{tabular}{c|ccc}
\hline \hline
LHC & 7 TeV & 8 TeV & 14 TeV \\
\hline
$\sigma_H(gg \to H)$ (pb) with $90\%$ C.L. errors&
 $13.4^{+4.8\%}_{-5.0\%}$ &
 $17.0^{+4.6\%}_{-4.6\%}$ &
 $44.5^{+5.2\%}_{-5.2\%}$\\
\hline
\ \ \ \ \ \ \ \ \ \ \ \ \ \ \ \ \ \ with 68\% C.L. errors&
 $13.4^{+2.9\%}_{-3.2\%}$ &
 $17.0^{+2.8\%}_{-2.9\%}$ &
 $44.5^{+3.4\%}_{-3.2\%}$\\
\hline
\hline
\end{tabular}
\end{center}
\vspace{-2ex}
\caption{\label{tbl:contour100}
Higgs boson production cross sections via gluon fusion channel  at the LHC, with 7, 8 and 14 TeV.
The combined PDF and $\alpha_s$ uncertainties
at the $90\%$ C.L.
have been calculated by the Lagrange multiplier method in the CT10H analysis.
The errors are expressed as the percentage of the central value.
}
\end{table}

\begin{figure}[H]
\begin{center}
\includegraphics[width=0.47\textwidth]{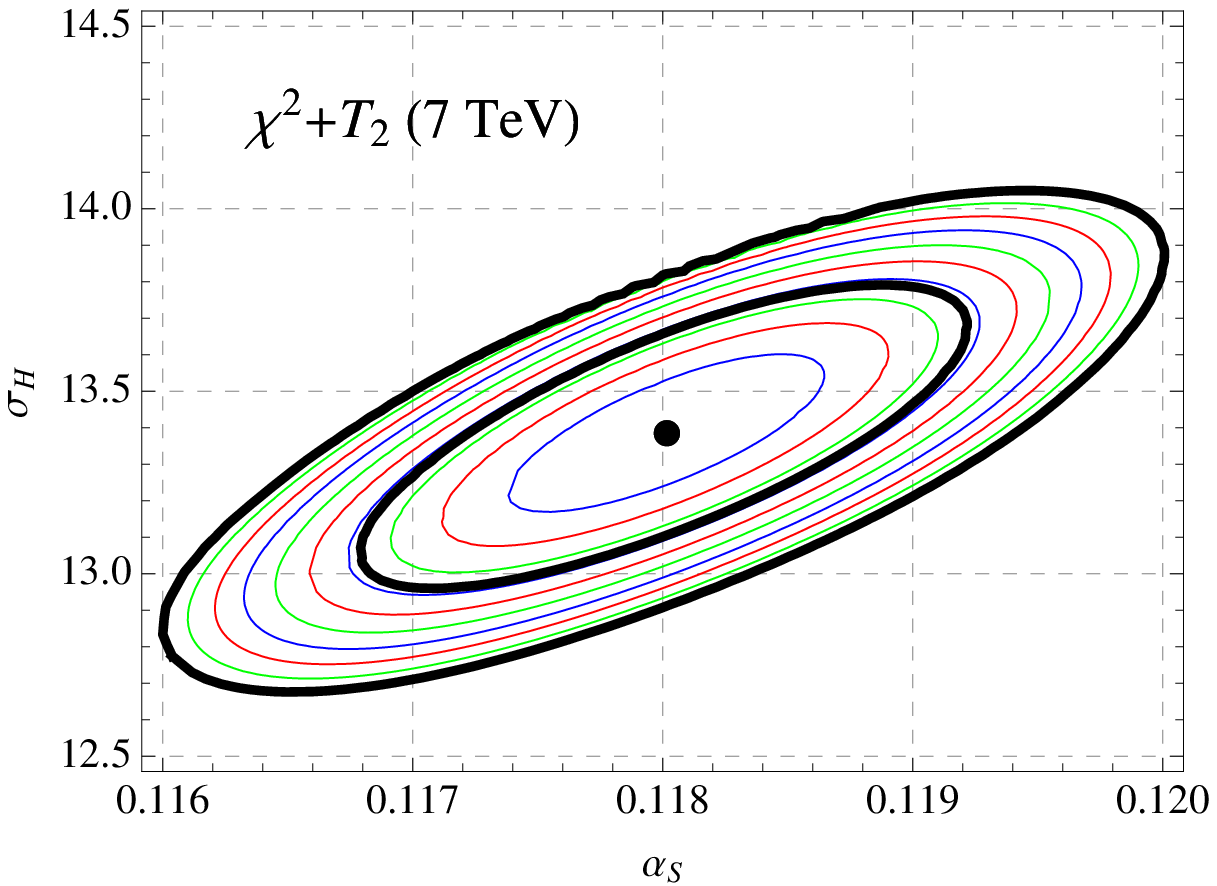} \,\,\,
\includegraphics[width=0.47\textwidth]{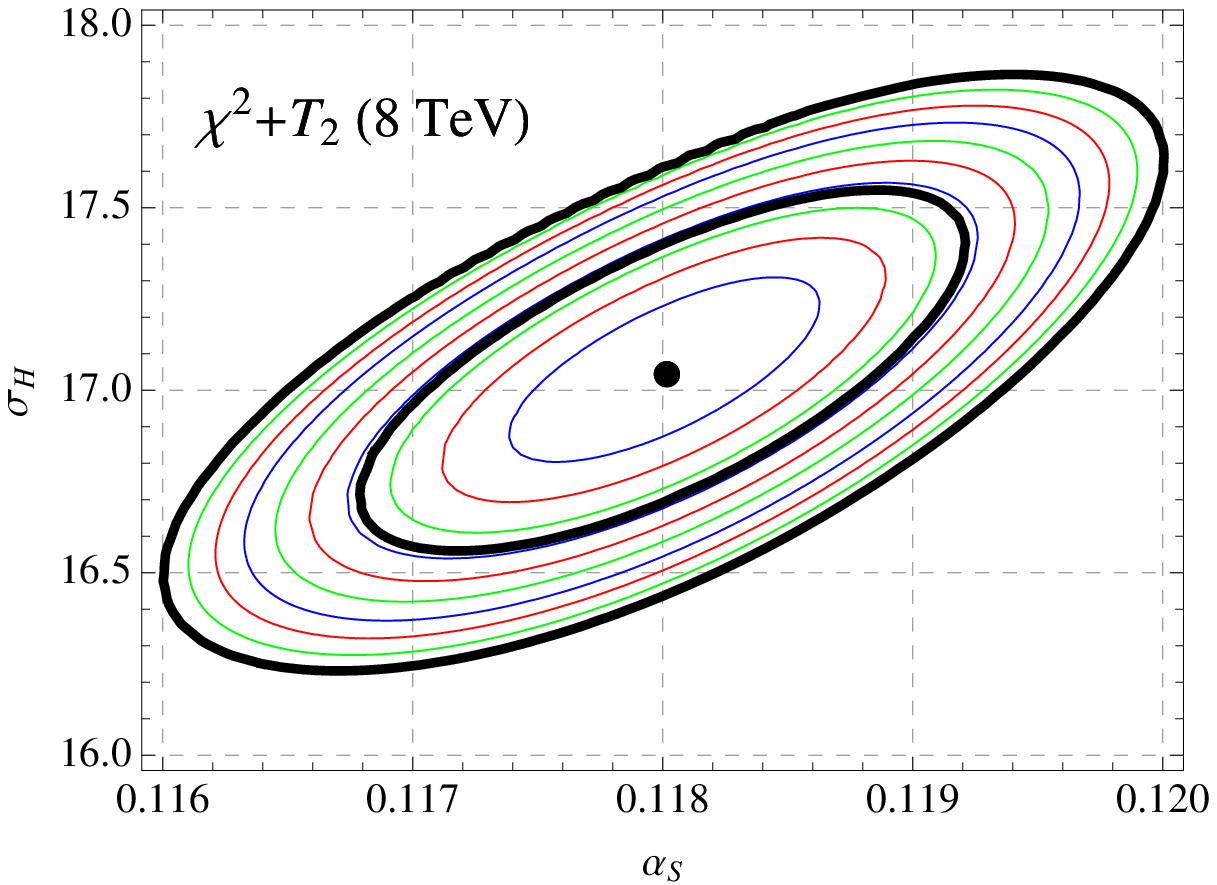}
\includegraphics[width=0.47\textwidth]{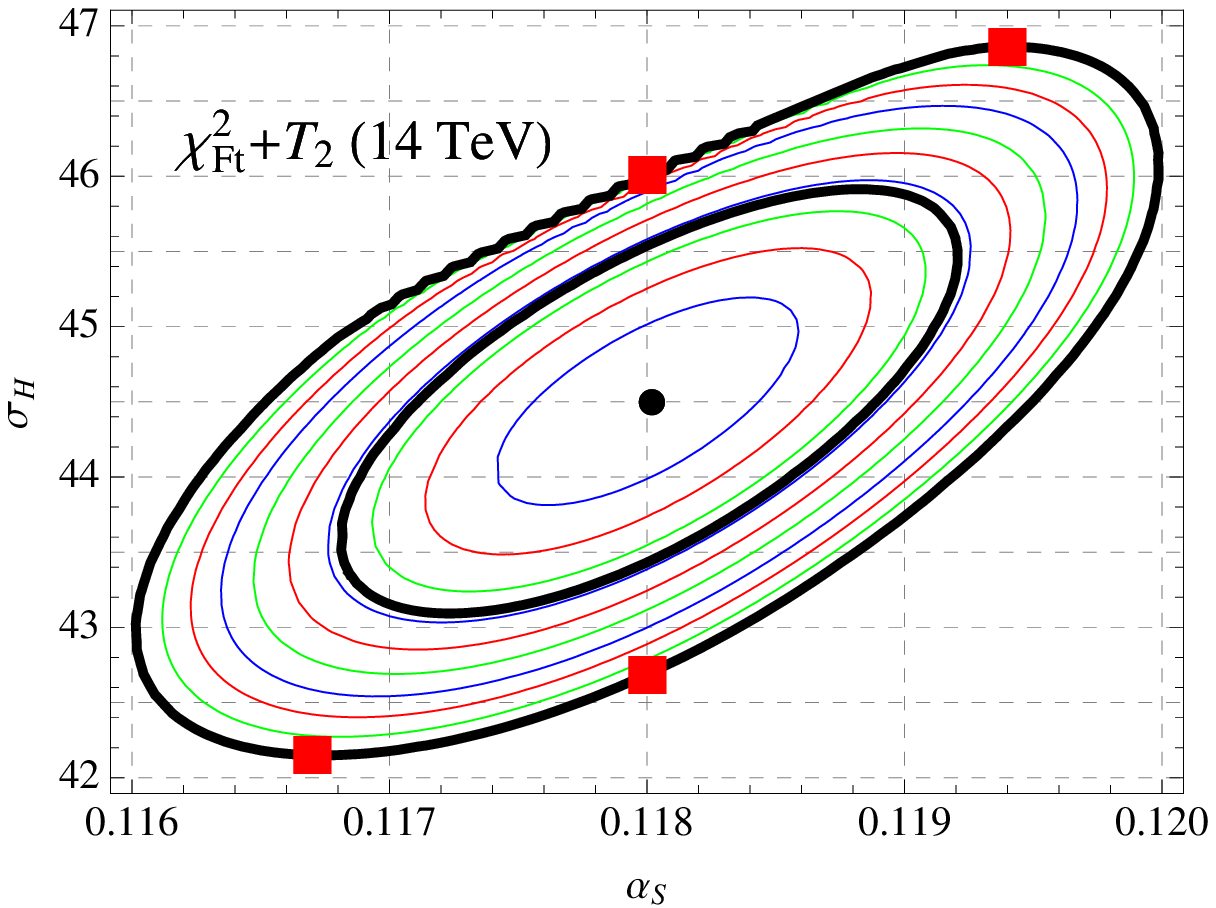}
\end{center}
\caption{
Contour plots of $\chi^{2}+\mbox{tier-2}$ ($T_2$)
in the $(\alpha_{s},\sigma_{H})$ plane,
for $\sigma_H$ (in pb unit) at the LHC, with 7, 8 and 14 TeV.
The thick black outer and inner contours are at
$\Delta\chi^2=100$ and $100/(1.645)^2$, respectively, for the $90\%$ C.L.
and $68\%$ C.L..  The thin colored contours are at intervals in $\chi^2$ of 10.
The fits that give minimum and maximum $\sigma_H$ are indicated by
the red square symbols, with $\alpha_s(M_Z)=0.1167$, $0.118$ or $0.1194$.
(See the text in Sec. IV.C for its details.)
\label{fig:contourswtier2}}
\end{figure}

Figure \ref{fig:contours} shows the minimum global $\chi^{2}$ value, without tier-2 penalty,
as a function of $(\alpha_{s},\sigma_{H})$.  However, we have argued previously that
including the tier-2 penalty with the $\chi^2$ function is a better indicator of goodness-of-fit.
Therefore, in Fig.~\ref{fig:contourswtier2}, we present contour plots of $\chi^{2} +$ tier-2 penalty
in the $(\alpha_{s},\sigma_{H})$ plane,
for $\sigma_H$ at the LHC with $\sqrt{s} = $ 7, 8 and 14 TeV.
The tier-2 penalty has a small effect for 7 and 8 TeV.
The effect is larger for 14 TeV,
especially for $\sigma_{H} \gg \sigma_{H}(a_0)$.
The tier-2 penalty does reduce the uncertainty of the prediction
of $\sigma_{H}$:
the area enclosed by any contour is smaller in Fig.~\ref{fig:contourswtier2}.
However, the reduction of uncertainty is fairly small.
In addition, the change in the maximum and minimum values of $\sigma_H$ along
the $\Delta\chi^2=100$ contour is negligible, even for $\sqrt{s}=14$ TeV.

\subsection{Comparisons of LM and Hessian uncertainties}

From Tables \ref{tbl:LMtable} and \ref{tbl:contour100} (LM method)
and Table \ref{tbl:xsecs} (Hessian method)
we can compare the PDF-only uncertainties, as well as the combined
PDF + $\alpha_s$ uncertainties, on the $gg\to H$ cross section computed
by the two methods.
The PDF-only uncertainties are for $\alpha_s(M_Z) = 0.118 $.
The central values of $\sigma_H(\,7\ {\rm TeV})$, $\sigma_H(\,8\ {\rm TeV})$
and $\sigma_H(\,14\ {\rm TeV})$ are identical in both calculations
by definition, so we can use the percent error to compare the uncertainties.
Both methods give asymmetric errors, which are compared in Table \ref{tbl:3errs}.

\begin{table}\begin{center}\begin{tabular}{l|ccc|ccc}
\hline\hline
&\multicolumn{3}{c|}{90\% C.L.}&\multicolumn{3}{c}{68\% C.L.}\\
\hline
Method & 7 TeV & 8 TeV & 14 TeV & 7 TeV & 8 TeV & 14 TeV \\
\hline
LM (PDF-only)                     &\ +3.2/-3.7\ &\ +3.2/-3.7\ &\ +3.5/-4.1\ &\ +2.0/-2.2\ &\ +2.0/-2.3\ &\ +2.2/-2.4\  \\
Hessian (PDF-only)                &\ +3.0/-3.0\ &\ +3.2/-3.1\ &\ +4.3/-3.6\ &\ +1.8/-1.8\ &\ +1.9/-1.9\ &\ +2.6/-2.2\ \\
LM (PDF + $\alpha_{s}$)      &\ +4.8/-5.0\ &\ +4.6/-4.6\ &\ +5.2/-5.2\ &\ +2.9/-3.2\ &\ +2.8/-2.9\ &\ +3.4/-3.2\ \\
Hessian (PDF + $\alpha_{s})$ &\ +4.7/-4.6\ &\ +4.8/-4.7\ &\ +5.4/-5.0\ &\ +2.9/-2.8\ &\ +2.9/-2.8\ &\ +3.3/-3.0\ \\
\hline
\hline
\end{tabular}\end{center}
\vspace{-2ex}
\caption{\label{tbl:3errs}
Uncertainties of $\sigma_{H}(gg \to H)$ computed by the LM method
and by the Hessian method, with tier-2 penalty included.
The $90\%$ and $68\%$ C.L. errors are given as percentage of the central value,
and the PDF-only uncertainties are for $\alpha_s=0.118$.
}
\end{table}

From Table \ref{tbl:3errs} we see that the PDF-only uncertainties
(for $\alpha_s(M_Z) = 0.118 $)
are fairly similar in both methods of calculation.
The LM method tends to be slightly more asymmetric, with larger
negative uncertainties, due to the slight non-quadratic
behavior of the $\chi^2$ function. The case of $\sqrt{s}=14$ TeV is interesting
because the direction of the asymmetry in the errors is opposite between
the LM and Hessian methods.  However, for all collider energies, the
difference in the error estimates between the two methods are considerably
smaller than the estimates themselves, and also smaller than the general theoretical
uncertainty in defining the $90\%$ C.L. errors (the choice of tolerance $T$,
for example.)

For the combined PDF + $\alpha_s$ errors, the agreement between the two methods
of calculation is also good.  In the LM method, the errors tend to be less asymmetric
when the $\alpha_s$ uncertainty is included, which brings the estimates from the two
methods into even better agreement.  For example, for $\sqrt{s}=8$ TeV and $\sqrt{s}=14$ TeV
the difference between the two PDF + $\alpha_s$ error estimates is less than $5\%$ of the error estimates
themselves.  This C.L.early shows that the Hessian method of combining the PDF and the $\alpha_s$
errors in quadrature is valid, and that the Hessian method gives a reliable error estimate in the case
of the $gg\to H$ cross section.

The fact that the Hessian and Lagrange multiplier estimates are in good agreement
for the Higgs boson cross section is because
the $\chi^2$ dependence on the fitting parameters $\{a\}$ is
mostly quadratic in the relevant tolerance range, and that $\sigma_H$ is predominantly
a linear function of the parameters in the same range.  Furthermore, the tier-2 penalty does
not have a very large effect here, only turning on near the edge of the uncertainty range.
Thus, the error estimates from the Hessian method are in good, though not perfect, agreement with
those from the LM method.  In addition, this
explains why the Hessian method of adding the $\alpha_s$ uncertainties in quadrature works
quite well, and why the prescription for obtaining the
$68\%$ C.L. errors from the $90\%$ C.L. errors
is reasonable.

We must emphasize here that these conC.L.usions apply only to the $gg\to H$
cross section with $M_H=125$ GeV.
In particular, the assumption that the observable $(\sigma_{H})$
depends linearly on the fitting parameters $\{a\}$ over the relevant range
might not be true for other observables.
Even in the case of the Higgs boson cross section from gluon fusion,
one might have expected larger nonlinear effects, since the cross section
depends strongly on both the gluon PDF $g(x,Q)$ and the value of $\alpha_s$.
For other observables, which may be more sensitive to other aspects of the PDFs,
the nonlinear effects may be greater.
This may especially be true if the observable is strongly correlated with a single
experimental data set in the global analysis,
which would lead to a large contribution from the tier-2 term.
In this case the LM and Hessian methods would give larger and more significant differences,
with the LM method giving the more reliable error estimate.
Thus, the LM method provides an important alternative
to the simpler Hessian method.

In conclusion, for $\sigma_{H}$ the Hessian and LM methods give consistent results;
and the tier-2 penalties have small effects.
We find this somewhat surprising for the following reasons:
(i) We use a rather large tolerance value, $T \lesssim 10$,
which one might expect to allow nonlinearities in the dependences
of $\chi^{2}(a)$ and $\sigma_{H}(a)$ on $\left\{ a \right\}$. Meanwhile,
the Hessian method is based on a linear error analysis.
Nevertheless, the final results are consistent with the LM method,
which does not rely on linearity.
(ii) The fact that the uncertainties are {\em asymmetric} shows that
nonlinearities do exist;
but again the Hessian treatment of the asymmetric errors is satisfactory.
(iii) Simply combining PDF error and $\alpha_{s}$ error in quadrature
in the Hessian method, gives results similar to the full $\sigma_{H}$
and $\alpha_{s}$ correlated uncertainties obtained in the LM method.

Are the above results surprising or not?
We would not know whether the Hessian method is reasonable,
without completing the LM calculations.
This is important to know,
because the Hessian method---using the LHAPDF library of error PDFs---is the
only method available for most studies of PDF uncertainty.
Furthermore, the LM calculations are interesting for another reason.
They allow the construction of the {\em contour plots},
which demonstrate very dramatically the correlations
between $\sigma_{H}$ and $\alpha_{s}$ uncertainties.

\section{Correlations between $\sigma_H$  and PDFs}

The Hessian and LM calculations are each better suited for elucidating different aspects of
how the PDFs influence the uncertainty in the Higgs boson cross section.
In this section we examine some of these details for the two methods in turn.

\subsection{Error Sets and Correlation Cosines in the Hessian Method}

\begin{figure}[H]
\begin{center}
\includegraphics[width=0.5\textwidth]{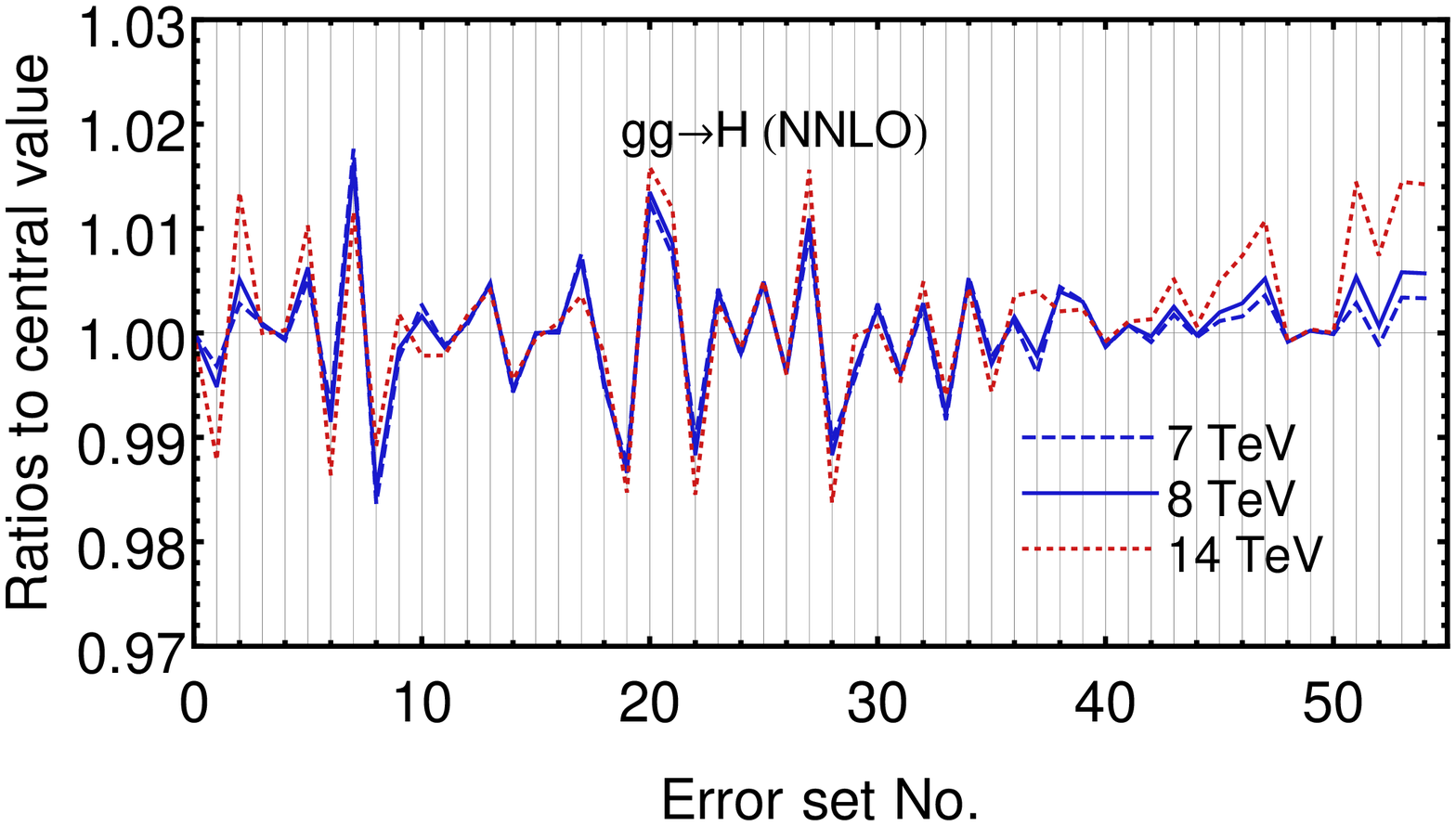}
\includegraphics[width=0.5\textwidth]{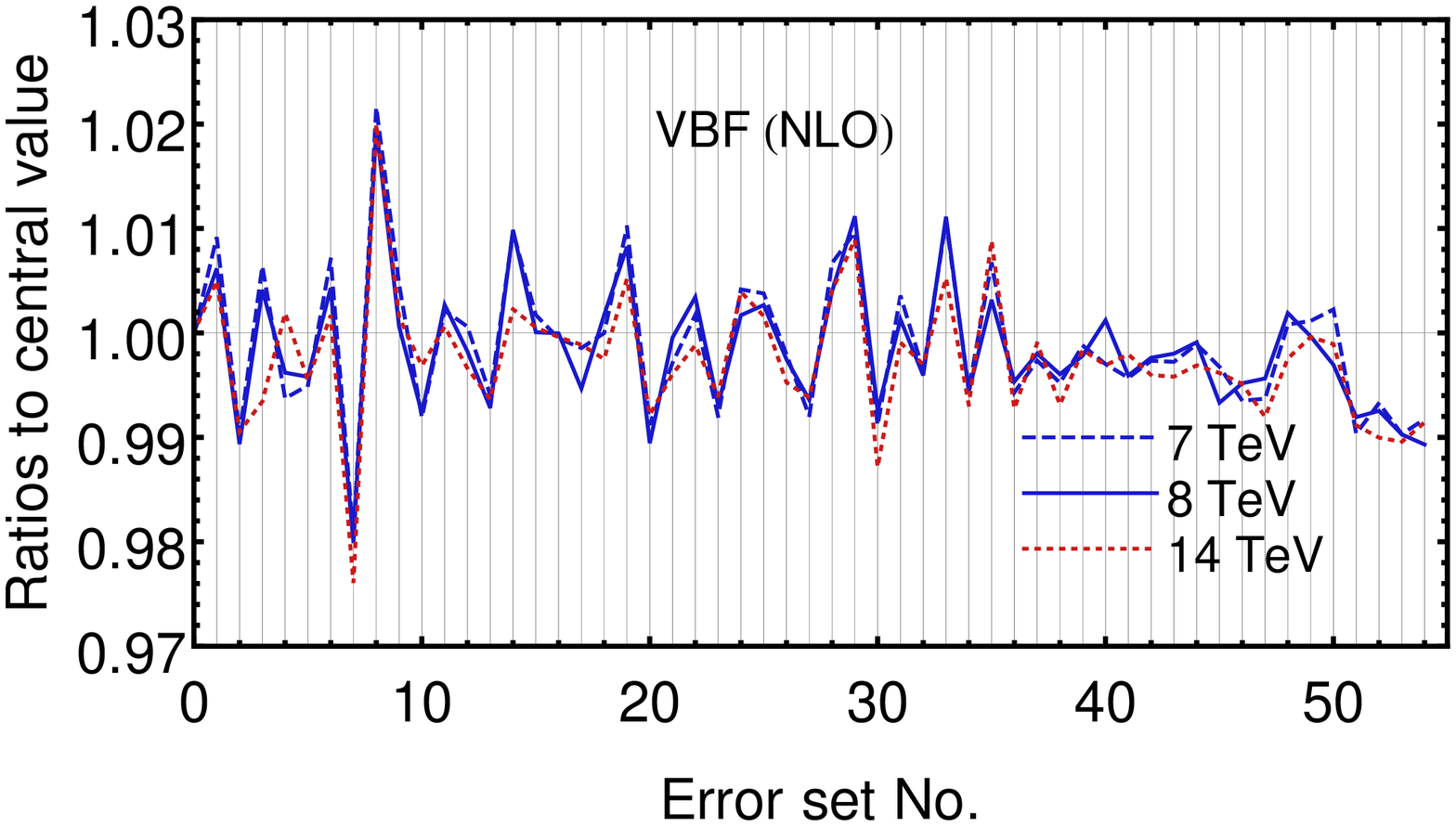}
\end{center}
\vspace{-5ex}
\caption{\label{fig:jun1a}
Ratio of the prediction of the Higgs boson cross section
from each error set to that from the central set.
The results for the $gg$ fusion and VBF channels are shown in
the upper and lower panels.  The dashed, solid, and dotted lines are for the LHC energy at 7, 8, and 14 TeV, respectively.
}
\end{figure}

The error sets that are obtained in the Hessian method can be used to compare the
sensitivity of different observables to the various PDF parameters~\cite{Nadolsky:2008zw}.
In Fig.~\ref{fig:jun1a} we plot the ratios of the predictions from each of the error sets to the best-fit set,
for the Higgs boson cross section at the LHC, in both the gluon fusion and vector boson fusion (VBF)
subprocesses.  In this study, we use a slightly enhanced set of PDF parameters, with two additional
eigenvectors in the PDF parameter space, as compared to the CT10 NNLO PDFs.
As usual, these error sets were obtained after
including the tier-2 penalty in the global analysis.
The VBF cross sections were calculated up to NLO using the
VBFNLO-2.6.1 code~\cite{vbfnlo}, with both the renormalization and the
factorization scales set to $\mu=M_H = 125\ {\rm GeV}$,
and with all the other default settings,
including a minimal invariant mass cut of $600\ {\rm GeV}$
for the two tagging jets.\footnote{The jet selection cuts
are $p_T>20\,{\rm GeV}$ and $|y|<4.5$, with the anti-$k_T$ jet
algorithm and a distance parameter $D=0.8$. Neither NLO electroweak
correction nor third jet veto is applied in the calculation.}

The results for the $gg$ fusion and VBF channels are shown in
the upper and lower panels of Fig.~\ref{fig:jun1a}. The dashed, solid, and dotted lines are for the LHC energy at 7, 8, and 14 TeV, respectively.
It is interesting to note  that
the relative importance of the major error sets
({\it i.e.}, the eigenvectors in the PDF parameter space)
is roughly independent of the collision energy.
In the case of the $gg\to H$ cross section,
we see that the PDF uncertainties
at all three energies are dominated by a few eigenvectors, which are associated with the
variations of the gluon PDF.
Furthermore, the values for the $gg$ fusion and for the VBF subprocesses in these figures tend to be opposite in
sign (at least for the first few error sets, with largest eigenvalues),
indicating the anti-correlation of the two subprocesses.
Namely, the error sets that increase the $gg \to H$ cross section
will decrease the VBF cross section, and vice versa.
Moreover, the PDF induced errors in the ratios of $gg \to H$ cross sections
at different LHC energies are expected to be small, about $2\%$
with its center value around 3.3 in the ratio of 14 TeV to 7 TeV predictions,
evaluated at the 90\% C.L.. A similar result holds for the VBF process, with its center value around 4.4.

\begin{figure}[H]
 \begin{center}
  \includegraphics[width=0.32\textwidth]{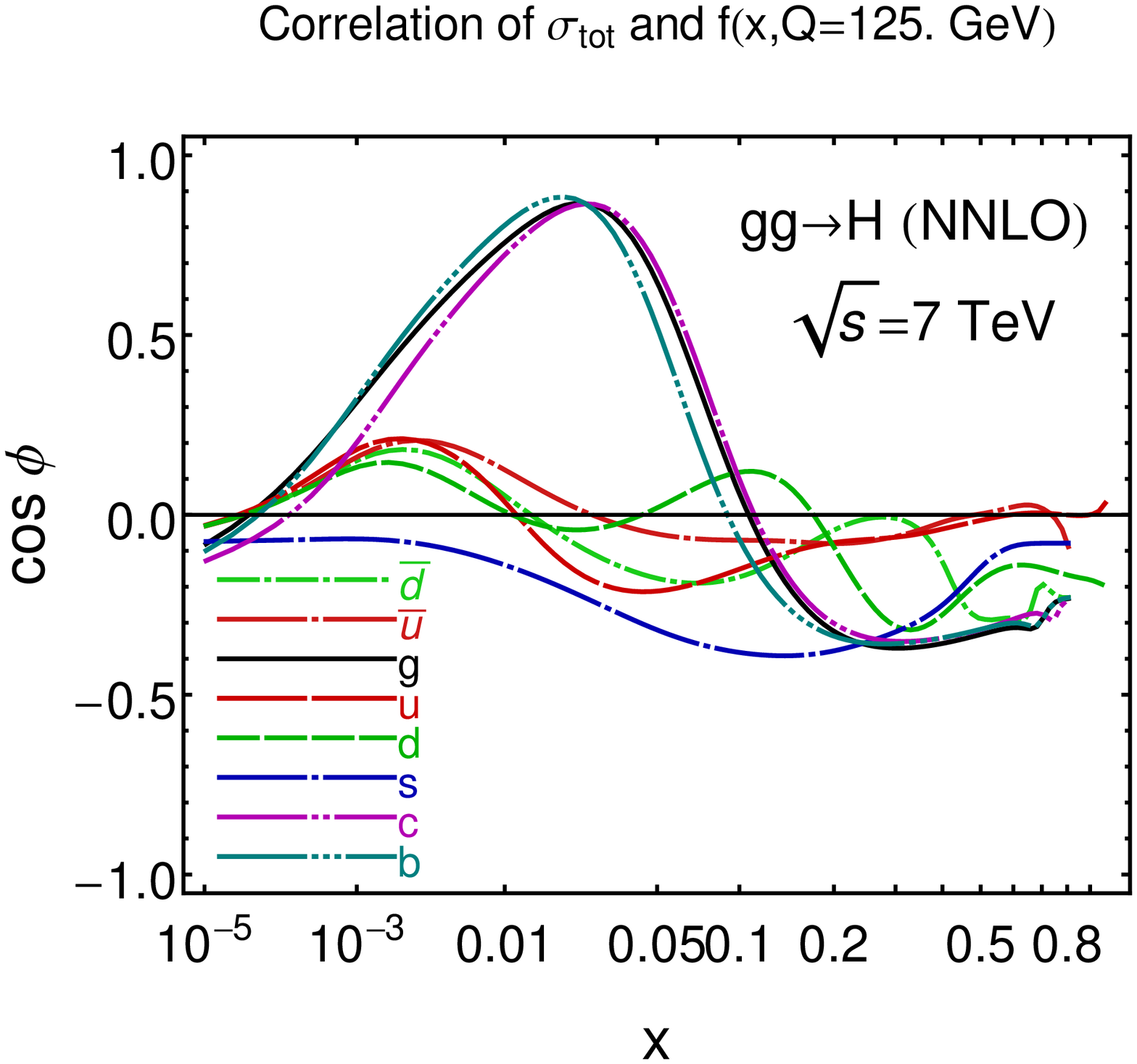}
  \includegraphics[width=0.32\textwidth]{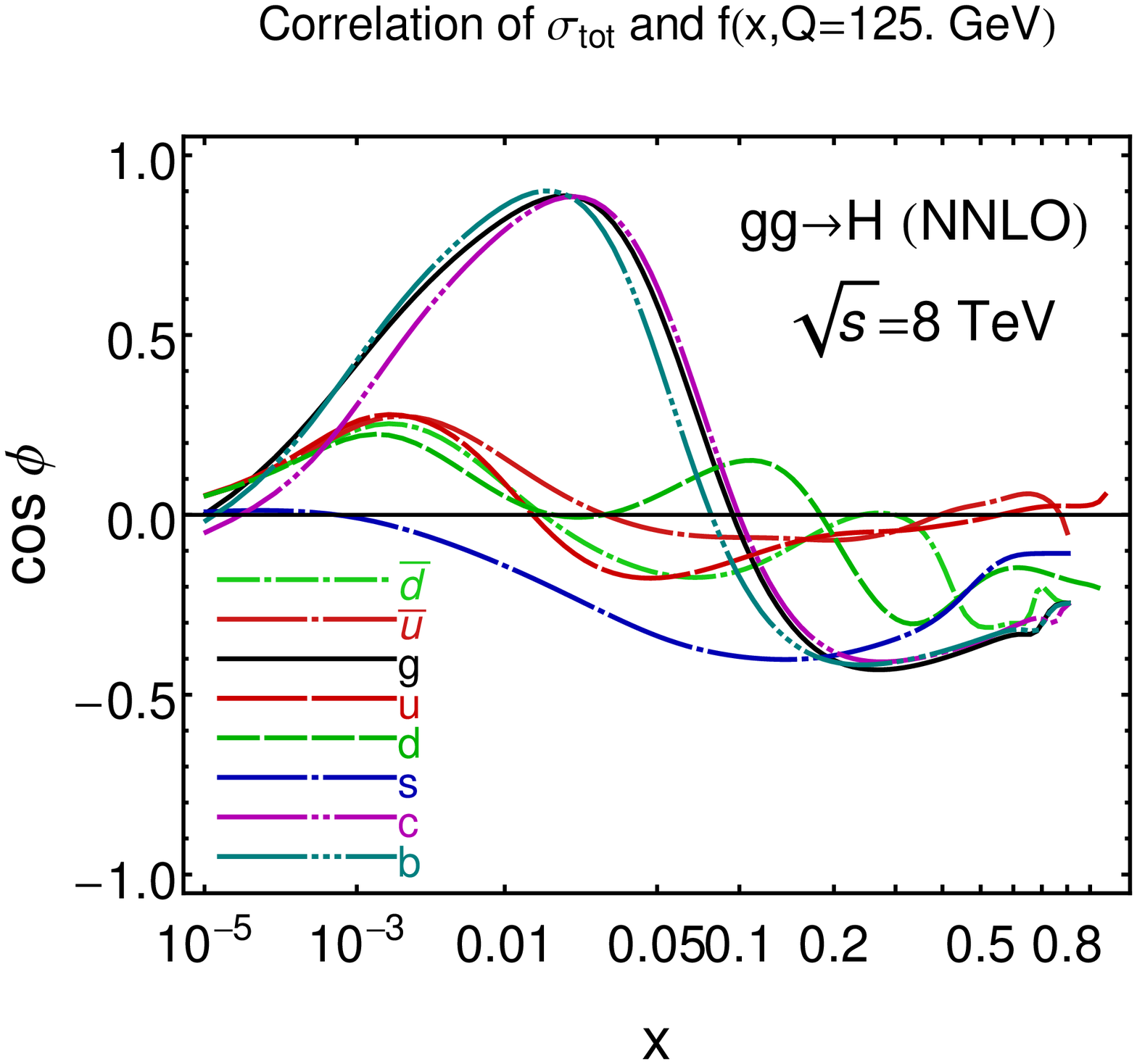}
  \includegraphics[width=0.32\textwidth]{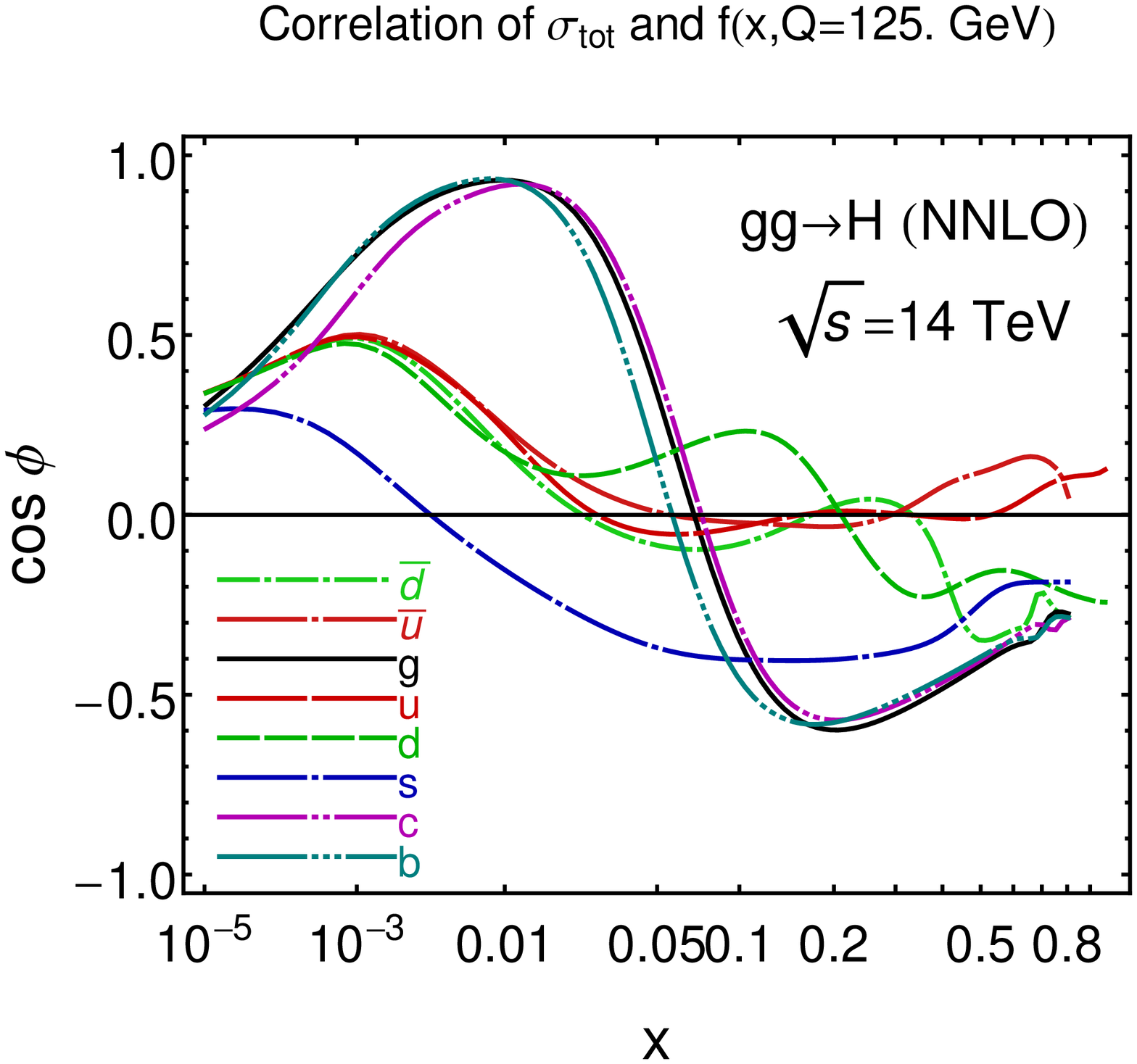}
 \end{center}
\vspace{-5ex}
\caption{\label{fig:jun2}
Correlation cosine between the $gg \to H$ cross sections
and the PDFs at $Q=125\ {\rm GeV}$ as functions of $x$,
at the LHC, with 7, 8, and 14 TeV.}
\end{figure}

\begin{figure}[H]
 \begin{center}
  \includegraphics[width=0.32\textwidth]{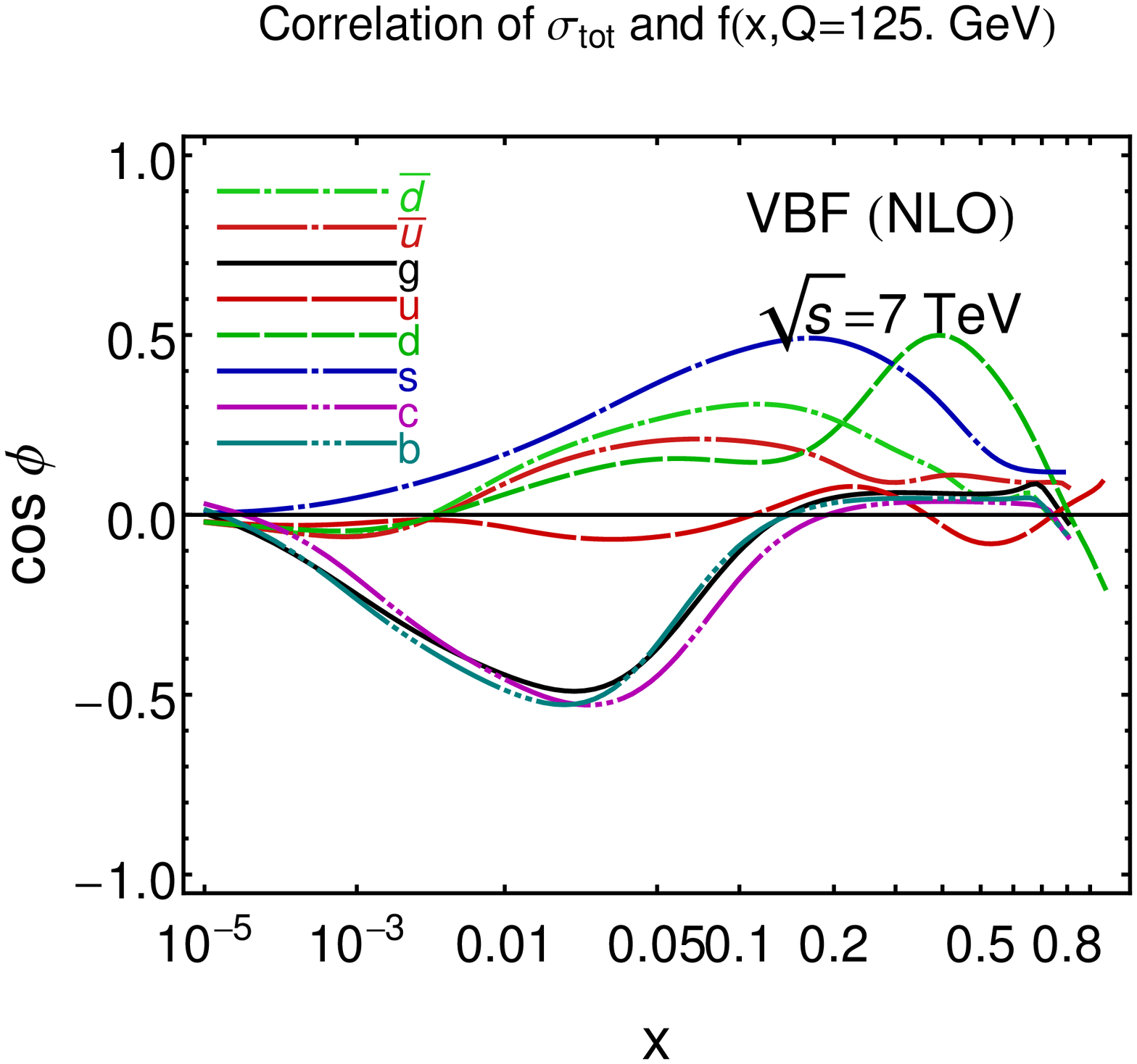}
  \includegraphics[width=0.32\textwidth]{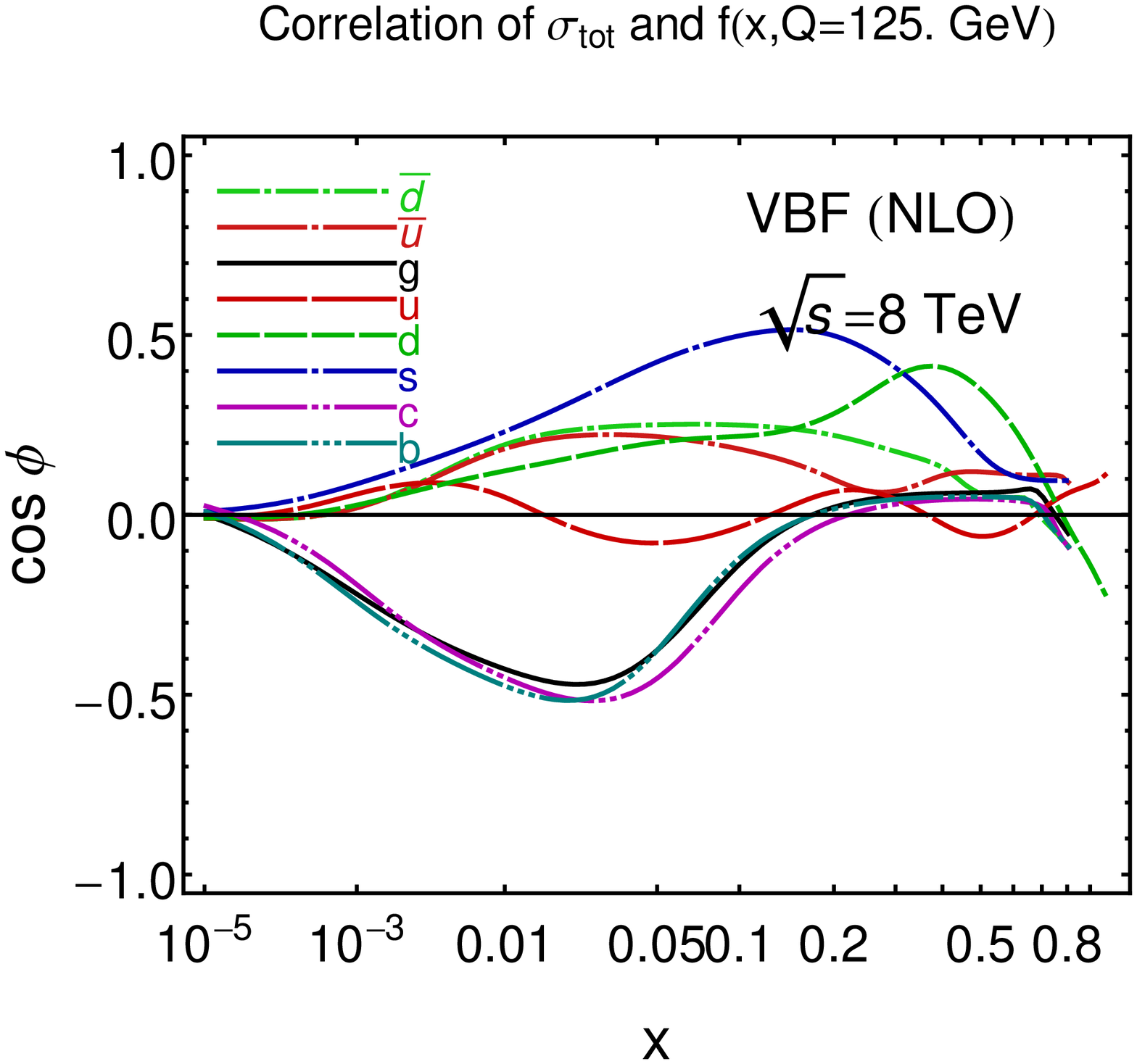}
  \includegraphics[width=0.32\textwidth]{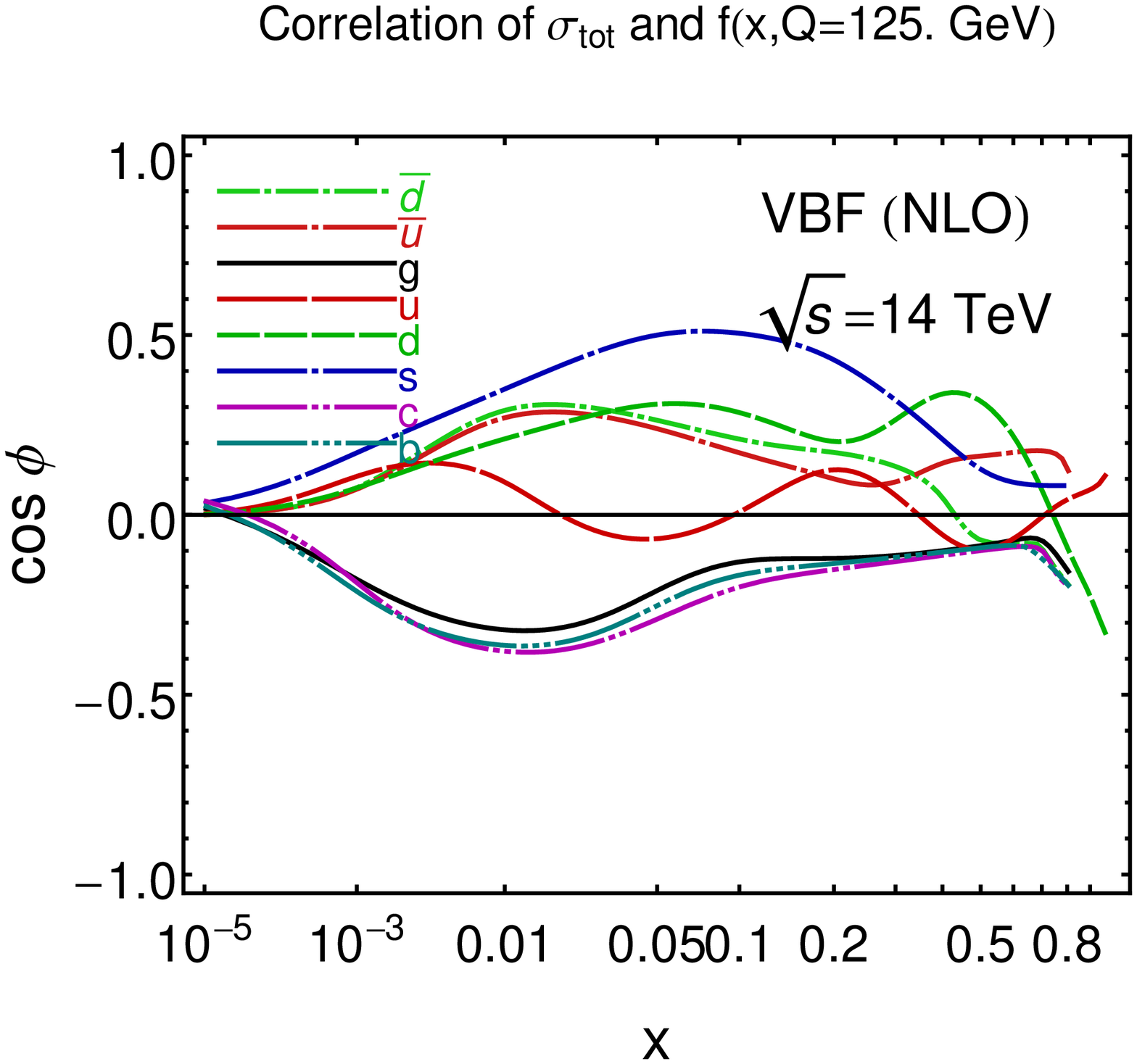}
 \end{center}
\vspace{-5ex}
\caption{\label{fig:jun3}
Correlation cosine between the VBF component of $\sigma_{H}$
and the PDFs at $Q=125\ {\rm GeV}$ as functions of $x$,
at the LHC, with 7, 8, and 14 TeV.}
\end{figure}

\begin{figure}[H]
 \begin{center}
  \includegraphics[width=0.32\textwidth]{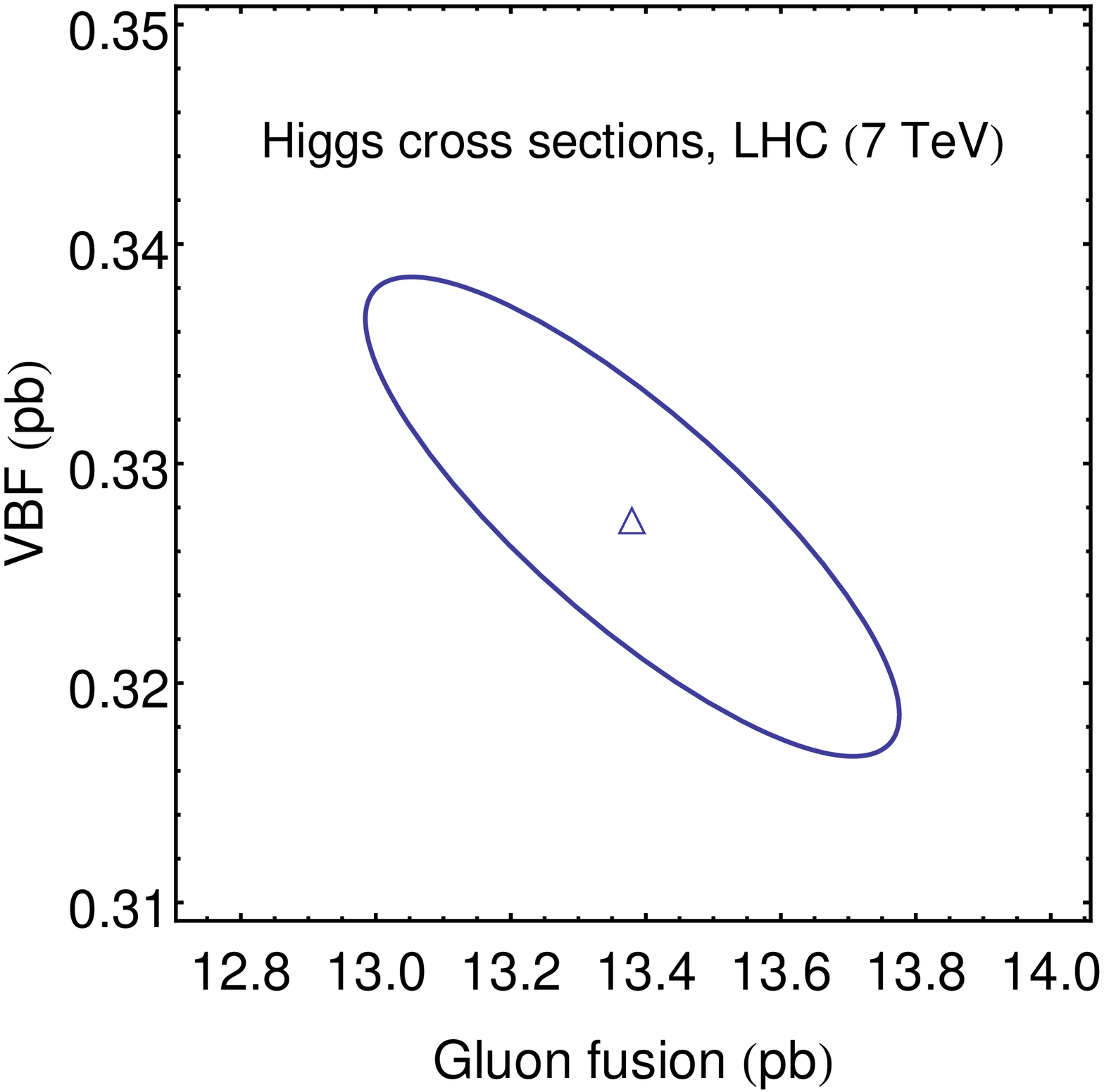}
  \includegraphics[width=0.32\textwidth]{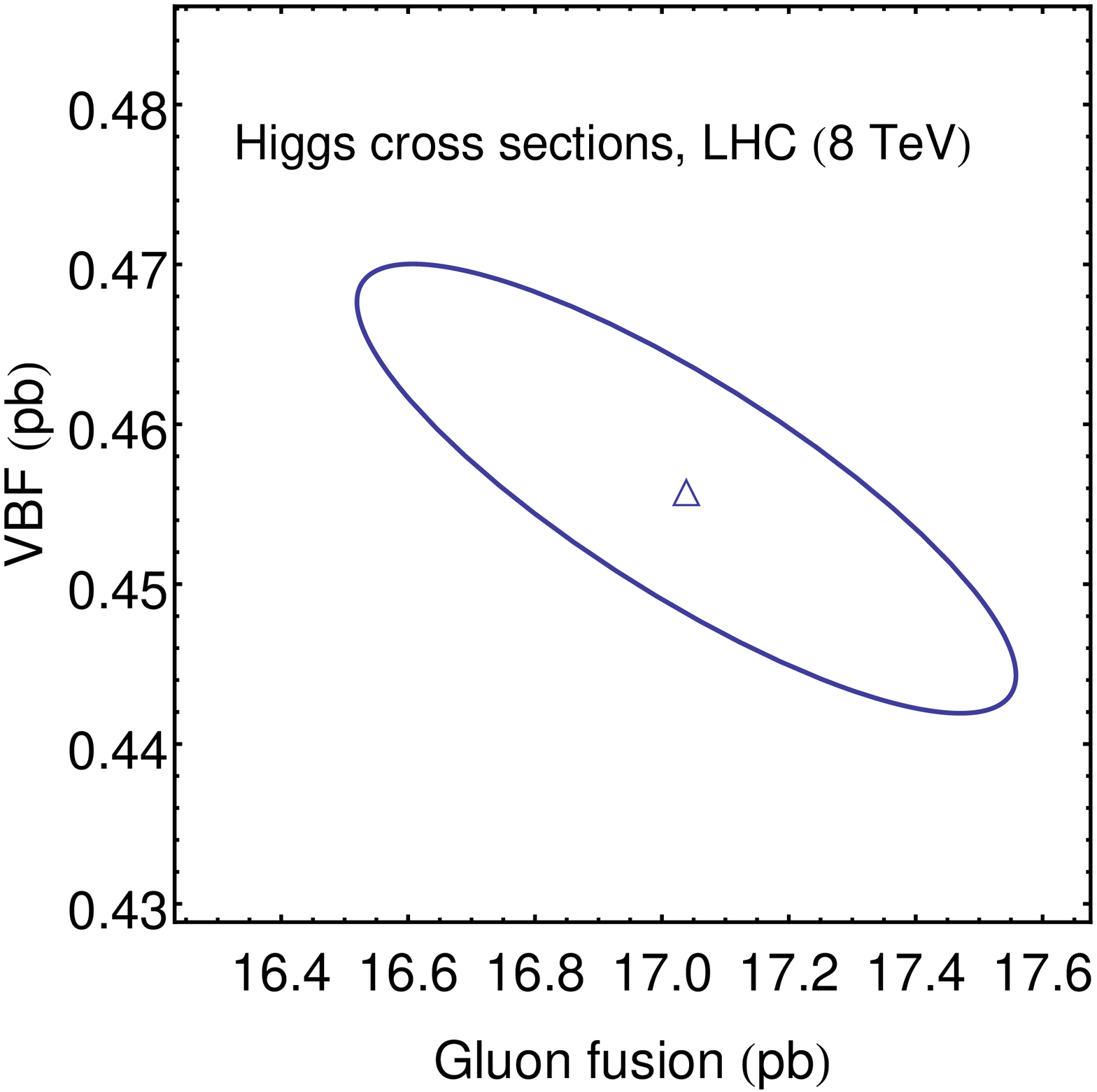}
  \includegraphics[width=0.32\textwidth]{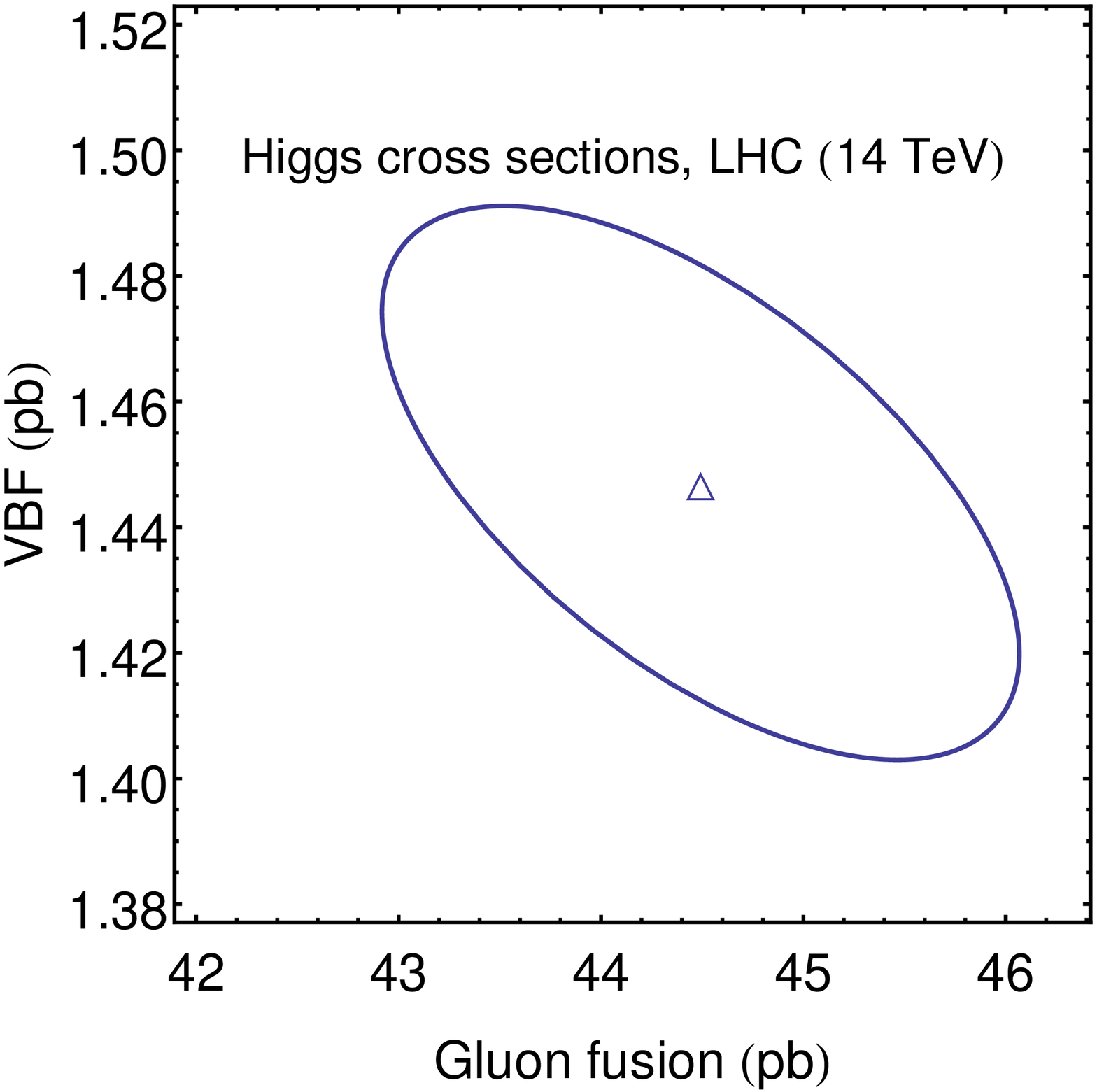}
 \end{center}
\vspace{-5ex}
\caption{\label{fig:jun4}
The $90\%$ tolerance ellipse of $(\sigma_{H})_{gg}$ and $(\sigma_{H})_{\rm VBF}$
at the LHC, with 7, 8, and 14 TeV.}
\end{figure}

In the Hessian approach, assuming quadratic approximation,
we can also study the direction of the gradient of the
Higgs boson cross section in the PDF parameter
space~\cite{Pumplin:2001ct,Nadolsky:2008zw, Nadolsky:2001yg}.
Figure \ref{fig:jun2} shows the correlation
between $\sigma_{H}$ and the PDFs of different flavors,
as a function of the parton momentum fraction $x$.
The correlation of two observables is measured by the cosine
of the angle between the gradient directions of the two observables
in the PDF parameter space~\cite{Nadolsky:2008zw}.
From Fig.~\ref{fig:jun2} we can see a strong correlation between
the $gg\to H$ cross section and the gluon PDF at $x \sim 0.01$,
as expected. The charm and bottom PDFs track the gluon PDF in these plots,
since they arise through gluon splitting.
Figure \ref{fig:jun3} shows a similar, but weaker, correlation with the gluon PDF
for the VBF process.   The
correlations between the gluon PDF and the two different subprocesses are opposite
in sign, consistent with the error PDF plots in Fig.~\ref{fig:jun1a}.
We can see this moderate anti-correlation directly
in the 90\% C.L. correlation ellipse of the two Higgs boson production subprocesses,
as shown in Fig.~\ref{fig:jun4}.

\subsection{Correlations between data sets and $\sigma_H$ in the LM method}
\label{sec:LMcorr}

Consistent with the error analysis in the LM method,
we can learn more about the impact of data on the PDF analysis
by calculating the {\em correlations} between individual data sets
and the PDFs with constrained values of $\sigma_{H}$.
In this section, we identity the experimental data sets that
correlate most strongly with the constrained value of $\sigma_H$
(via the $gg$ channel) from the Lagrange multiplier calculations.

In a LM calculation, the constrained value of $\sigma_{H}$ can be
pushed to a larger or smaller value, as compared to its central
value $\sigma_H(a_0)$ (corresponding to $\lambda=0$),
by changing the value of $\lambda$; cf. Eq.~(\ref{eq:lmf}).
We now examine the degree to which the change in $\sigma_H$ causes
a specific data set to agree less well with the
theory prediction than for $\lambda = 0$.
We need a measure of goodness-of-fit to address the question.
We could compare variations of the figures-of-merit
$\chi_{e}^{2}$ for each experimental data set
in the scan over $\lambda$, but would find such comparison
inscrutable because of different sizes $N_{pt}$ of the experimental
data sets and, consequently, their incompatible $\chi^2_e(N_{pt})$
distributions. Instead, we make use
of an `equivalent Gaussian variable'' $S$ \cite{Dulat:2013hea},
introduced originally as a part of the tier-2 penalty \cite{Lai:2010vv}.

For each particular data set assumed to obey a chi-squared
probability distribution, we map the $\chi^2_e({a},N_{\mathrm pt})$
value of the data set onto a respective variable $S_{e}$,
which has the same cumulative probability,
but obeys a Gaussian distribution with unit standard deviation.
A more detailed definition of $S_e$ and its relation to the $\chi^2$-probability
distribution function can be found in Ref.~\cite{Dulat:2013hea}.
A value of $S_e$ in the range of $-1$ to $1$ indicates a
good agreement (at the $68\%$ C.L.) between data and theory,
analogous to $\chi_{e}^{2}\approx N_{\mathrm pt}$, in the case of large $N_{\mathrm pt}$.
$S_e$ much larger than $+2$ indicates a poor fit,
analogous to $\chi_{e}^{2} \gg N_{\mathrm pt}$.
$S_e$ much less than $-2$ indicates an unexpectedly good fit,
much better than one would expect from normal statistical analysis;
{\it i.e.}, they have anomalously small residuals,
presumably because the true experimental errors are smaller than
the published values.

\begin{figure}[H]
\begin{center}
\includegraphics[width=0.47\textwidth]{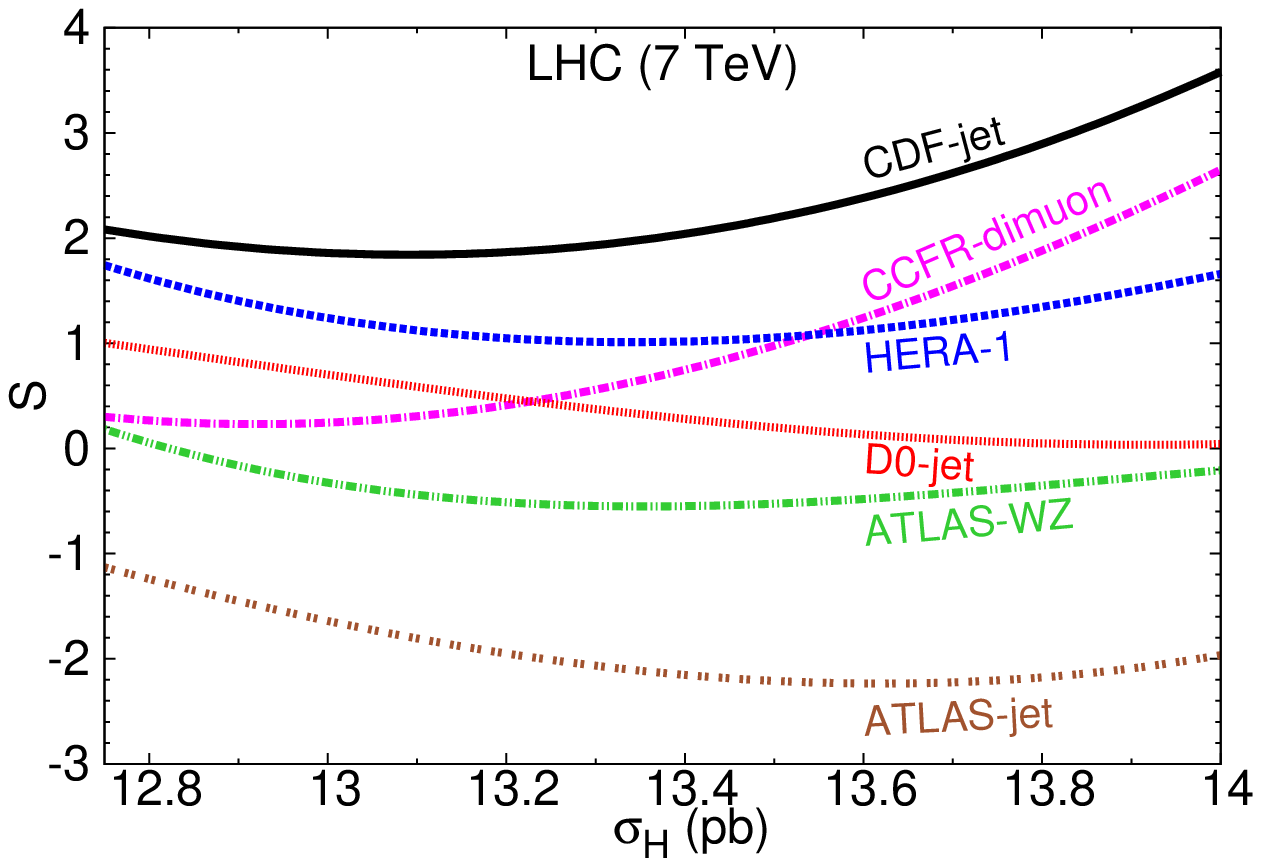}
\includegraphics[width=0.47\textwidth]{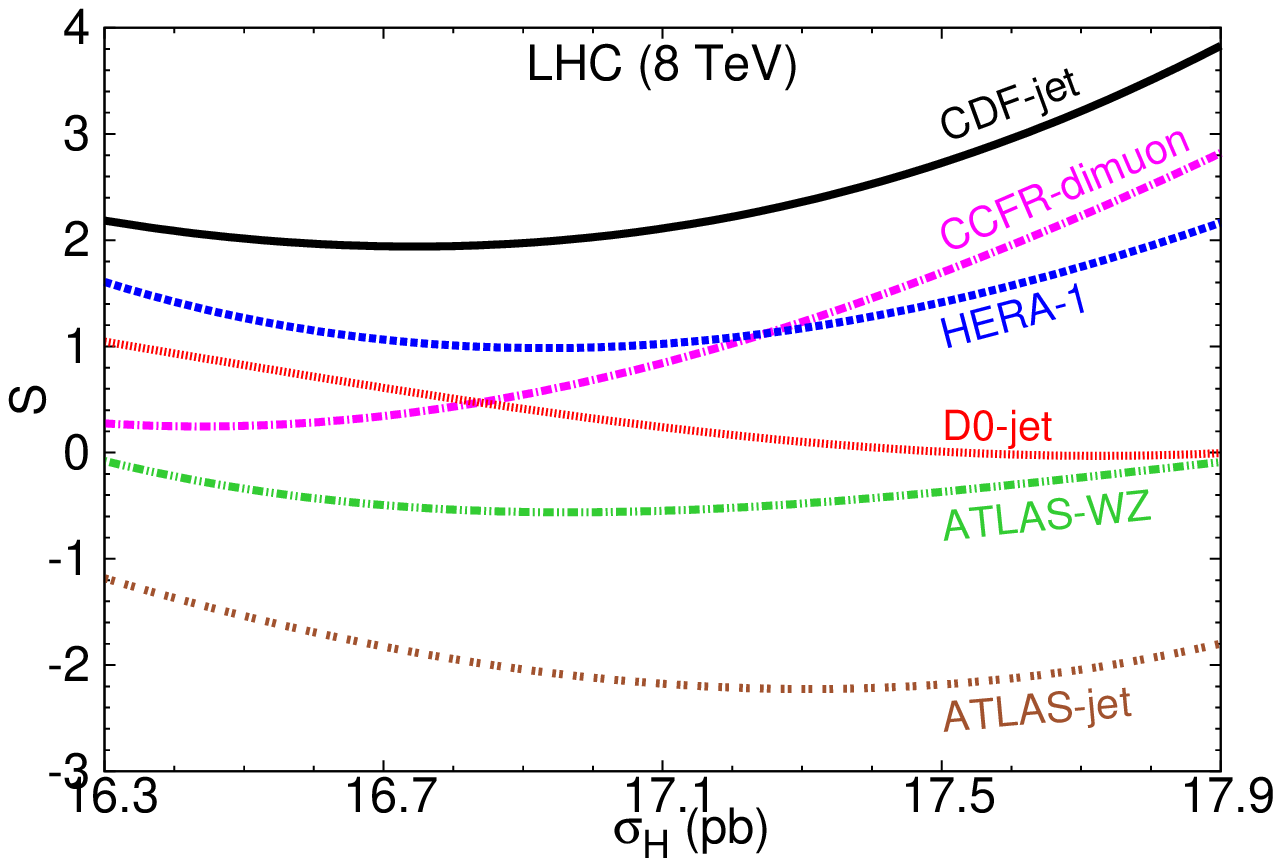}
\includegraphics[width=0.47\textwidth]{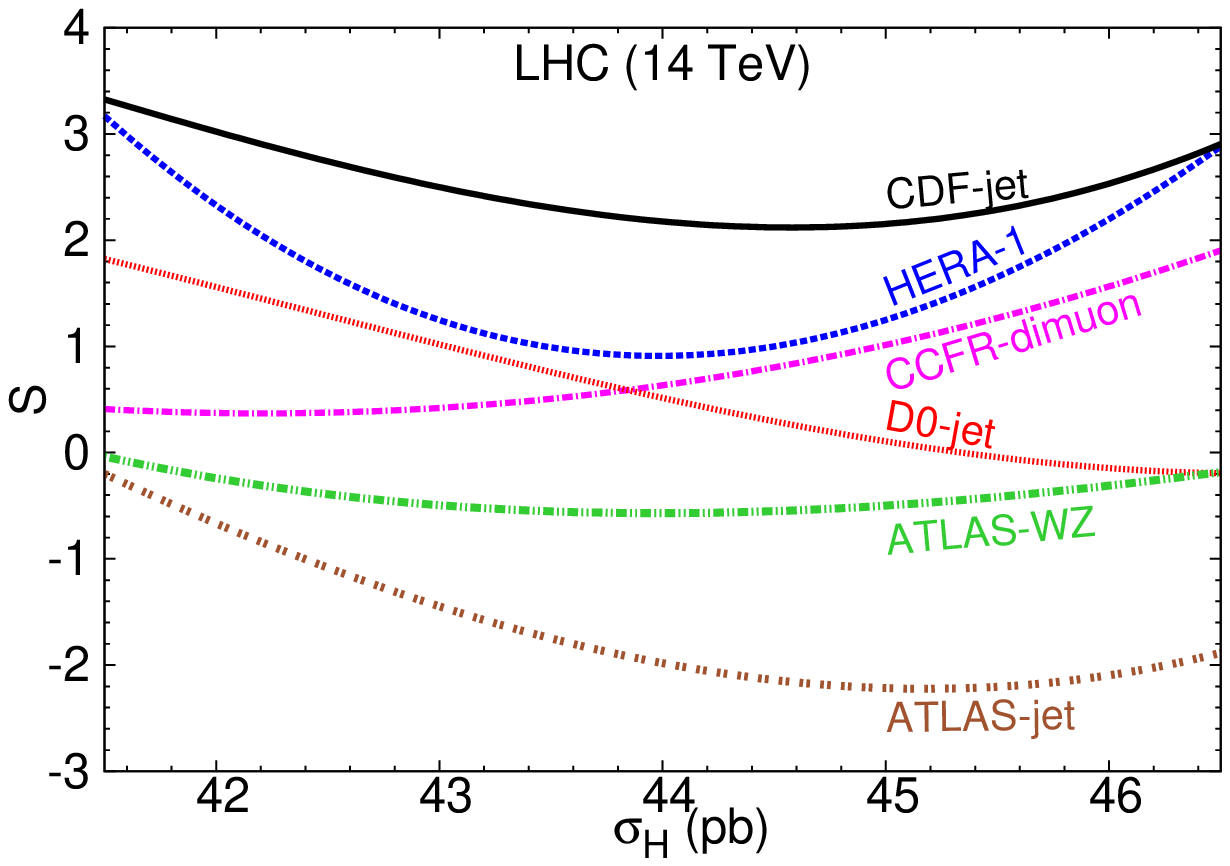}
\end{center}
\caption{The equivalent Gaussian variable $S$ versus $\sigma_H$ (in pb unit)
at the LHC, with 7, 8 and 14 TeV.
The labels CDF, D0, HERA-1, CCFR-dimuon, ATLAS$_{\rm WZ}$, and ATLAS$_{\rm jet}$ correspond to
the experiment ID numbers 504, 514, 159, 126, 268, and 535, respectively, given
in Table \ref{tab:EXP_bin_ID}.
\label{fig:spartan}}
\end{figure}

Something important is gained by this mapping.
The statistical meaning of the value
of $\chi^{2}_e/N_{\mathrm pt}$ depends on $N_{\mathrm pt}$;
but the mapping to $S_{e}$ removes this complication. For example, the
confidence levels on $S_e$ in the previous paragraph are independent of
$N_{pt}$, so in Figure \ref{fig:spartan} we plot $S_e$ to
compare sensitivities of the data sets to the changes in the PDFs
that also change  the prediction for $\sigma_{H}$.
A similar figure showing $\chi^{2}/N_{\mathrm pt}$ would be far less informative:
the meaning of each curve would depend on $N_{\mathrm pt}$,
which varies from 8 to 579 among the 33 data sets (cf.~Table 1).

In Fig.~\ref{fig:spartan}, the values of $S$ are shown versus the
constrained values of $\sigma_H$ at the LHC with 7, 8, and 14 TeV
for the four experimental data sets with ID numbers 126, 159, 504 and 514
(cf. Table~\ref{tab:EXP_bin_ID}).
These are the experiments whose $S$ values show strong correlation to the
constrained values of $\sigma_H$, at all three energies of the LHC.
We also plot the values of $S$ for the two LHC data sets,
ID numbers 268 and 535.
As we push $\sigma_{H}$ away from its central value $\sigma_{H}(a_0)$,
we see that $S$ increases substantially for some data sets,
corresponding to a worse agreement between data and theory.
The upper four experiments shown in Fig.~\ref{fig:spartan} are sensitive to
the constrained theoretical value of $\sigma_{H}$,
because the increase in $S$ signals a decrease in
likelihood for that value of $\sigma_H$.

It is not surprising that the $S$ dependence indicates that
the inclusive high-$p_T$ jet production measurements
at the Tevatron by the CDF
(ID number 504) and D0 (ID number 514) Collaborations
are strongly correlated
to the constrained value of $\sigma_H$ at the LHC,
because these are the experimental data that are
sensitive to the gluon PDF.
Similarly, the marked variation of $S$ for the
combined HERA Run 1 data set (ID number 159) is understood,
because it is well known that the
HERA data provide important constraints on determining the
gluon PDF.
The pattern of sensitivity to $\sigma_{H}$ is similar
for all three LHC energies, although we note that the
combined HERA Run 1 data set
increases in importance as the collider energy increases and smaller $x$ values
for the gluon are sampled.
The sensitivity of $S$ for the
CCFR neutrino dimuon data \cite{Goncharov:2001qe}
(ID number 126)
is less obvious. After a careful examination, we find that
in the CT10H PDF sets, when the gluon PDF increases to push up
the constrained value of $\sigma_H$, the strange PDF decreases
at a $Q$ value around 3 GeV and $x$ value of order $10^{-2}$,
so as to strongly reduce the predictions compared to the
CCFR dimuon data in that kinematic region. This is mainly due to
the momentum sum rule imposed on the parton distribution functions.

A final observation is that the
$S$ values for the two LHC data sets, the combined
ATLAS $W^\pm$ and $Z$ data (ID number 268) and
the ATLAS inclusive jet data (ID number 535), are all
smaller than zero within the $90\%$ C.L. range of
$\sigma_H$, at all three energies of the LHC, as seen in Fig.~\ref{fig:spartan}.
This indicates that these data are easily fit in the global analysis,
and hence, they
do not play a significant role in constraining the Higgs boson
cross section at the LHC, as predicted by the CT10H PDF sets.
Needless to say, these conclusions could change in the future
with more precise data from the LHC.

\subsection{Extreme PDFs from the Lagrange multiplier fits}

To facilitate the study of PDF uncertainties in the prediction
of $gg \to H$ total cross section,
we make available on LHAPDF~\cite{lhapdf} a few PDFs implementing the
findings of our CT10H Lagrange multiplier study.  The CT10H ensemble
consists of the central set, two sets for determination of the 90\%
C.L. PDF uncertainty on $\sigma_H$
at 14 TeV with a fixed $\alpha_s(M_Z)=0.118$
(corresponding to the red square symbols in
Fig.~\ref{fig:contourswtier2}), and two other sets to determine the
90\% C.L. extremes on $\sigma_H$ from the combined PDF+$\alpha_s$ analysis
at 14 TeV.  In the second pair of PDF sets, the
upper and lower uncertainty limits on $\sigma_H$ correspond
to $\alpha_s(M_Z)=0.1194$ and $0.1167$, respectively.
The CT10 and CT10H central sets are entirely compatible and can be
used interchangeably. While the extreme sets are derived
to predict the PDF-induced errors in $\sigma_H$
at the LHC with $14 \, \mathrm{TeV}$, they
also reproduce the corresponding errors at 7 and 8 TeV
to sufficient accuracy.

\begin{figure}[H]
\begin{center}
\includegraphics[width=0.47\textwidth]
{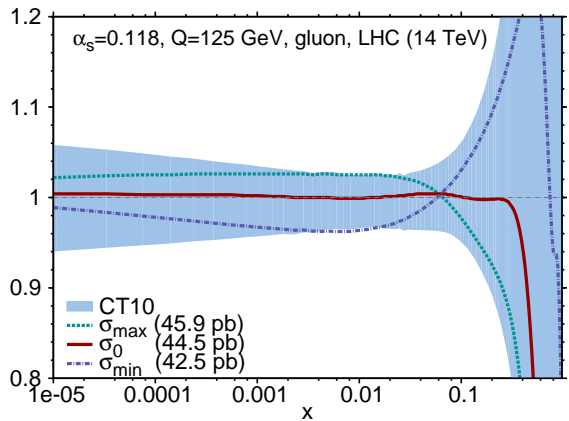}
\end{center}
\caption{
CT10H gluon PDFs at the momentum scale 125 GeV, compared to the
CT10 error band, at the $90\%$ C.L..  These CT10H fits give the
central prediction ($\sigma_0$),
and the minimum ($\sigma_{min}$) and maximum
($\sigma_{max}$) predictions obtained using the
Lagrange multiplier method, for $\sigma_H$ at the LHC with 14 TeV, as listed in Table~\ref{tbl:LMtable}. Also, $\alpha_s(M_Z)=0.118$.
\label{fig:LMgluon}}
\end{figure}

Figure \ref{fig:LMgluon} compares the gluon
distributions from CT10H to the CT10 NNLO uncertainty band,
with $\alpha_s = 0.118$, at the $90\%$ C.L..
One sees that the CT10H uncertainty is practically the same as the
CT10 NNLO range for $0.002 \lesssim x \lesssim 0.03\,$, which
is the region dominating the Higgs boson total cross section
via gluon fusion channel at the LHC. In this $x$ region, the pair of
CT10H PDFs significantly reduces the computational burden in
estimating the uncertainties, compared to CT10.
The CT10H sets underestimate the PDF-induced uncertainty
in predicting the kinematical distributions of the
Higgs boson sensitive to the gluon PDF with $x$ less than $10^{-4}$
or above $0.05$, where the full CT10 PDF ensemble is needed.
The CT10H sets can be also used to estimate the PDF uncertainty in
processes that are {\it strongly} correlated with $gg\rightarrow H$
production. For instance, the CT10H extreme sets predict more
than a half of the CT10 PDF-induced error in
Higgs boson production cross section via vector boson fusion,
because of the moderate anti-correlation of the $gg \to H$ and
VBF processes discussed in Sec. IV.A, and about a half of the CT10
error in the associated  $t {\bar t} H$ production.

Our general-purpose PDFs at this time remain CT10 NNLO PDFs.
The CT10H extremes should only be used for calculations of uncertainties
specifically related to Higgs boson production at central rapidities
at the LHC. In the future, when additional LHC data become available,
the CTEQ-TEA collaboration will construct a new generation of
general PDFs including that data in the analysis.

\section{Conclusions} \label{sec:conclusion}

Accurate predictions for the rates of Higgs boson production
are crucial for precision tests of the Higgs mechanism.
Studying the production rates and decay branching ratios of the Higgs boson
can potentially discriminate between models of electroweak symmetry
breaking, and the goal is to measure them
with an accuracy better than $10\%$~\cite{Dittmaier:2011ti,SnowmassHiggs2013}.

In the dominant gluon-gluon fusion channel, the two key contributions to
the error in the Higgs boson production rate are the uncertainties of the PDFs and the
QCD coupling $\alpha_s$.
In this paper, we addressed the estimation of the PDF uncertainty,
as well as the combined PDF+$\alpha_s$ uncertainty, for the
Higgs boson production cross section at the LHC.
In general, various methods for estimation of the PDF uncertainty
may yield somewhat different results,
with potential phenomenological implications.
For example, nonlinearities in the log-likelihood function $\chi^2$ employed in
the PDF fits could produce differences between
the error estimates obtained by different methods of the analysis.
Thus, it is important to determine the magnitude of the difference.

In the present work, we have used the CT10 NNLO global analysis
(with minor updates), which we call CT10H NNLO, to investigate this issue.
We compared two
standard methods for performing the error analysis: the Hessian method,
which utilizes an eigenvector decomposition of the PDF parameter space;
and the Lagrange multiplier  method, which utilizes constrained
global fits of the PDFs.  A set of eigenvector PDFs has been
previously distributed by the CT10 NNLO global
analysis~\cite{ct10nnlo} to compute the PDF uncertainty for any
experimental observable in the Hessian method.
By comparing the results from this method to those from the LM method,
which makes fewer assumptions
about the functional dependence of $\chi^2$ on the PDF parameters, we have
checked the reliability of the Hessian result in the particular case of the
production of SM Higgs boson, with a 125 GeV mass,
via the gluon-gluon fusion channel.

Our main conclusion from this analysis is that the two methods give
quite similar predictions for the PDF uncertainty on the $gg\to H$
cross section, thus supporting previous results that relied
exclusively on the Hessian method.  For the pure PDF
uncertainty, both methods give relative errors of about
$\pm3$ to $\pm4\%$ at the 90\% C.L. for the different energies,
with small differences between the methods.  For the combined
PDF+$\alpha_s$ error, the two methods agree even better, validating the
prescription of adding the Hessian PDF errors and the
$\alpha_s$ errors (obtained from the $\alpha_s$ series of
PDFs) in quadrature.  For example, the
combined uncertainty at $\sqrt{s}=14$ TeV was found to be $\pm5.2$\% in the LM
method and $+5.4/-5.0$\% in the Hessian method at the 90\% C.L..
Both methods in general yield a mild asymmetry in the errors.
The differences between the two methods are
certainly less than other theoretical uncertainties,
such as in the choice of tolerance used to define the 90\% C.L..
We have tried several nonperturbative parametrization
forms for the CT10H PDFs, but found little change in the predictions
and no change in our overall conclusions.

The agreement between the two methods is not trivial, given the
variety of involved factors. It can be traced to the fact that the
$\chi^2$ dependence on $\sigma_H$ is close to parabolic and is not
strongly affected by constraints from individual
experiments ({\it i.e.}, by the tier-2 penalty), within the tolerance range.
In addition, the approximately quadratic behavior in $\chi^2$ implies
that the method for
obtaining a 68\% C.L. interval from the 90\% C.L. interval, by dividing by 1.645,
is appropriate.
We emphasize that these fortunate features of the $\chi^2$ function may not
hold for other processes that are sensitive to other PDF flavors or to
the gluon PDF in a different range of $x$. A full comparison of two methods
would need to be repeated to check if they agree for any other observable.
Furthermore, if the observable is strongly correlated with a single experimental
data set in the global analysis, which would lead to a large contribution from
the tier-2 penalty, the LM and Hessian methods would yield different
predictions, with the LM method giving a more reliable error estimate.

In this paper we also presented some details of the uncertainty analysis that were
obtained with the two methods.  Using the Hessian method, we showed that the PDF
dependence produces a strong anti-correlation between the prediction for the rate of
Higgs boson production
through gluon-gluon fusion and that for vector boson fusion, at all three collider energies.
Using the LM method, we investigated the correlation between the value of $\alpha_s(M_Z)$
and the Higgs boson cross section in gluon-gluon fusion, as displayed in the contour plots of
Figs.~\ref{fig:contours} and \ref{fig:contourswtier2}.  We also checked the impact of individual experimental data sets on the prediction of the Higgs boson cross section.
As expected, the data sets that constrain $\sigma_H$ most strongly
are the Tevatron jet data and the HERA Run 1 DIS data,
which are most sensitive to the gluon PDF.  Agreement with the HERA data set, in particular, is strongly sensitive
to variations in the Higgs boson cross section at 14 TeV.

From the LM scan we have obtained a pair of PDF sets that are sufficient to determine the two
uncertainty extremes in the $gg\to H$ cross section at $\sqrt{s}=14$ TeV,
with a fixed $\alpha_s(M_Z)=0.118$.
Similarly, we have obtained a pair of PDF sets that determine the uncertainty extremes
from the combined PDF+$\alpha_s$ analysis at 14 TeV.
In this second pair of PDF sets, the one that gives the upper uncertainty limit
on $\sigma_H$ corresponds to a value of $\alpha_s(M_Z)=0.1194$,
while the one that gives the lower uncertainty limit corresponds
to a value of $\alpha_s(M_Z)=0.1167$.
Although the equivalent error sets for other energies will be slightly different,
we have checked that both of these pairs of PDF sets reproduce the corresponding
errors found at 7 and 8 TeV to reasonable accuracy.
These PDF sets will simplify the experimental analysis on the uncertainty in the
Higgs boson total cross section through gluon-gluon fusion,
compared to the standard Hessian method analysis,
as only two PDF error sets are needed to compute the PDF error.
These PDF sets are to be included in the LHAPDF library~\cite{lhapdf}
and also be made available via the internet website~\cite{ct10web}
which hosts all CT10 PDFs.

\clearpage

\section*{Appendix 1.
Analytic study of $\alpha_s$ and $\sigma_H$ uncertainties
\label{sec:app2}}

In the formula for $\chi^2$ with the $\kappa$-penalty, given by Eqs.~(\ref{eq:chi2})
and (\ref{eq:kappa}),
we can consider the fit parameters as functions of $\sigma_H$ and $\alpha_s\equiv\alpha_s(M_Z)$,
({\it i.e.,} $a_i\equiv a_i(\sigma_H,\alpha_s)$),
implicitly via the Lagrange multiplier calculation.
In this way we can treat $\chi^2$ as a function of $\sigma_H$ and $\alpha_s$:
\begin{eqnarray}
\label{eq:chitwo}
\chi^2(\sigma_H,\alpha_s)&\equiv&\chi^2(a_i(\sigma_H,\alpha_s),\alpha_s)\nonumber\\
&=&\chi_0^2(\sigma_H,\alpha_s)+\kappa\left(\frac{\alpha_s-\bar\alpha}{\delta\bar\alpha}\right)^2
\end{eqnarray}
where $\chi_0^2(\sigma_H,\alpha_s)$ is the $\chi^2$ value for $\kappa=0$.
In this appendix, we will consider this formula in the quadratic approximation
in order to understand the interplay of the PDF and $\alpha_s$ contributions
to the uncertainty on the Higgs boson cross section.  Consequently, we shall relate these results to the combined
uncertainty obtained in the Hessian method.

For $\kappa=0$, and working in the quadratic approximation,
we can write
\begin{eqnarray}
\label{eq:kzero}
\chi_0^2(\sigma_H,\alpha_s)&\approx&\chi_0^2(\sigma^0_H,\alpha^0_S)
+M_{11}(\sigma_H-\sigma_H^0)^2+M_{22}(\alpha_s-\alpha_s^0)^2\nonumber\\
&&\qquad\qquad\quad
+\,2M_{12}(\sigma_H-\sigma_H^0)(\alpha_s-\alpha_s^0)\ ,
\end{eqnarray}
where $\sigma_H^0$ and $\alpha_s^0$
are the best-fit values for $\kappa=0$, and
\begin{eqnarray}
M_{11}&=&\frac{1}{2}\frac{\partial^2\chi_0^2}{\partial {\sigma_H}^2}\Bigg|_{\sigma_H^0,\alpha_s^0}\, \nonumber\\
M_{22}&=&\frac{1}{2}\frac{\partial^2\chi_0^2}{\partial {\alpha_s}^2}\Bigg|_{\sigma_H^0,\alpha_s^0}\, \\
M_{12}&=&\frac{1}{2}\frac{\partial^2\chi_0^2}{\partial \sigma_H\partial \alpha_s}\Bigg|_{\sigma_H^0,\alpha_s^0}\ .\nonumber
\end{eqnarray}
Note that we are treating $\alpha_s$ as a fitting parameter.
Thus, $\alpha_s^0\equiv\alpha_{\rm GA}$ is the best-fit value as determined purely by the global analysis (GA) data.  From Eq.~(\ref{eq:kzero}) we can also obtain the 90\% confidence level uncertainty on $\alpha_s$
coming purely from the global analysis:
\begin{eqnarray}
\delta{\alpha}_{\rm GA}&=&\frac{T^2 M_{11}}{D}\ ,\label{eq:errorGA}
\end{eqnarray}
where $T$ is the tolerance level on $\chi^2$ used in the global analysis and
$D=M_{11}M_{22}-(M_{12})^2$ is the determinant of the matrix $M_{ij}$.  If we assume that the determination of
$\bar\alpha$ and $\alpha_{\rm GA}$ are uncorrelated, dependent upon different experimental constraints, we can obtain
the total combined uncertainty from
\begin{eqnarray}
\frac{1}{(\delta\alpha_{\rm WA})^2}&=&\frac{1}{(\delta\alpha_{\rm GA})^2}+\frac{1}{(\delta\bar\alpha)^2}\ ,\label{eq:errortot}
\end{eqnarray}
where we have identified the combined uncertainty with the world-average (WA) uncertainty, $\delta\alpha_{\rm WA}$.

For nonzero $\kappa$, we can write
\begin{eqnarray}
\label{eq:chiwithpenalty}
\chi^2({\sigma}_H,{\alpha}_S)&\approx&\chi^2(\sigma^\kappa_H,\alpha^\kappa_S)
+{M}^\kappa_{11}({\sigma}_H-\sigma^\kappa_H)^2+{M}_{22}^\kappa({\alpha}_S-\alpha^\kappa_S)^2\\
&&\qquad\qquad\quad
+2\,{M}^\kappa_{12}({\sigma}_H-\sigma^\kappa_H)({\alpha}_S-\alpha^\kappa_S)\ ,\nonumber
\end{eqnarray}
where $\sigma^\kappa_H$ and $\alpha^\kappa_S$
are the best-fit values with non-zero $\kappa$.
Note that all of the parameters in this equation can be obtained
as functions of the parameters in Eq.~(\ref{eq:kzero}), plus $\kappa$ itself.
We obtain the relations by plugging Eq.~(\ref{eq:kzero}) into Eq.~(\ref{eq:chitwo}),
and then comparing with Eq.~(\ref{eq:chiwithpenalty}).
The quadratic coefficients are
\begin{eqnarray}
{M}^\kappa_{11}&=&M_{11}\nonumber\\
{M}^\kappa_{22}&=&M_{22}+\kappa/(\delta\bar{\alpha})^2\\
{M}^\kappa_{12}&=&M_{12}\ .\nonumber
\end{eqnarray}

The best-fit values $\sigma^\kappa_H$, $\alpha^\kappa_S$, and $\chi^2(\sigma_H^\kappa,\alpha_s^\kappa)$
can be expressed in very intuitive forms, if we set $\kappa=T^2$ (the tolerance-squared used in the global analysis).
Then we find that the best-fit value for $\alpha_s$, for non-zero $\kappa$ is
\begin{eqnarray}
\label{eq:alphashifts}
\alpha_s^\kappa&=&(\delta\alpha_{\rm WA})^2\left[
\frac{\alpha_{\rm GA}}{(\delta\alpha_{\rm GA})^2} +
\frac{\bar\alpha}{(\delta\bar\alpha)^2}\right]\ .
\end{eqnarray}
Note that this is just the average of $\alpha_{\rm GA}$ and $\bar\alpha$, weighted by their relative variances.
We can identify $\alpha_s^\kappa\equiv\alpha_{\rm WA}$ as the world-average value, which incorporates all of the
experimental constraints on the strong coupling.
Similarly, we find
\begin{eqnarray}
\label{eq:sigmashifts}
\sigma_H^\kappa&=&(\delta\alpha_{\rm WA})^2\left[
\frac{\sigma_H(\alpha_{\rm GA})}{(\delta\alpha_{\rm GA})^2} +
\frac{\sigma_H(\bar\alpha)}{(\delta\bar\alpha)^2}\right]\ ,
\end{eqnarray}
where
\begin{eqnarray}
\sigma_H(\bar\alpha)\ \approx\ \sigma_H^0-(M_{12}/M_{11})(\bar\alpha-\alpha_s^0)\label{eq:sigmafixed}
\end{eqnarray}
is the best-fit result for $\sigma_H$ with fixed $\alpha_s=\bar\alpha$ (in the quadratic approximation),
and we have identified $\sigma_H^0\equiv\sigma_H(\alpha_{\rm GA})$.
Finally, we obtain a similar result for the minimum value of $\chi^2$ at non-zero $\kappa$.  We obtain
\begin{eqnarray}
\label{eq:kmin}
\chi^2(\sigma_H^\kappa,\alpha_s^\kappa)
&=&(\delta\alpha_{\rm WA})^2\left[
\frac{\chi_0^2(\sigma_H(\alpha_{\rm GA}),\alpha_{\rm GA})}{(\delta\alpha_{\rm GA})^2} +
\frac{\chi_0^2(\sigma_H(\bar\alpha),\bar\alpha)}{(\delta\bar\alpha)^2}\right]\ ,
\end{eqnarray}
where the global minimum with fixed $\alpha_s=\bar\alpha$ is given by
\begin{eqnarray}
\label{eq:kfixed}
\chi_0^2(\sigma_H(\bar\alpha),\bar\alpha)
&\approx&\chi_0^2(\sigma_H^0,\alpha_s^0)+(D/M_{11})(\bar\alpha-\alpha_s^0)^2\ ,
\end{eqnarray}
in the quadratic approximation, and we have identified $\chi_0^2(\sigma_H^0,\alpha_s^0)\equiv\chi_0^2(\sigma_H(\alpha_{\rm GA}),\alpha_{\rm GA})$.

Next we consider the combined uncertainty in $\sigma_H$ obtained from the general $\chi^2$ function
(\ref{eq:chitwo}) with a nonzero $\kappa$.  In the quadratic approximation,
it is straightforward to find the maximum and minimum values of $\sigma_H$ that
are consistent with a given $\Delta\chi^2$.
If we require
\begin{equation}
\chi^2(\sigma_H,\alpha_s)
-\chi^2(\sigma_H^\kappa,\alpha_s^\kappa)\le T^2\ ,
\end{equation}
 then we obtain
\begin{eqnarray}
({\sigma}_H-\sigma_H^\kappa)^2&\,\le\,&\left(\frac{{M}^\kappa_{22}}{{M}^\kappa_{11}{M}^\kappa_{22}-({M}^{\kappa}_{12})^2}\right)T^2\nonumber\\
&\,\le\, & \frac{T^2}{M_{11}}\ +\ \left(\frac{M_{12}\,\delta\alpha_{\rm WA}}{M_{11}}\right)^2\frac{T^2}{\kappa}
\label{eq:siguncertainty}\\
&\,\le\,&\Sigma_1\ +\ \Sigma_2\ ,\nonumber
\end{eqnarray}
where we have used Eqs.~(\ref{eq:errorGA}) and (\ref{eq:errortot}) to simplify the expression.
This gives the contribution to the uncertainty in the cross section as the sum of two terms in quadrature, each
of which has a specific interpretation.  The first term,
\begin{eqnarray}
\Sigma_1&=&\frac{T^2}{M_{11}}\ ,\label{eq:sig1}
\end{eqnarray}
is just the uncertainty-squared in the Higgs boson cross section due to the PDFs at fixed $\alpha_s$.
We can interpret the second term if we set $\kappa$ equal to our tolerance-squared ($T^2/\kappa=1$).  Then
\begin{eqnarray}
\Sigma_2&=&\biggl[\sigma_H(\alpha_{s}+\delta\alpha_{\rm WA})-\sigma_H(\alpha_{s})\biggr]^2\ ,\label{eq:sig2}
\end{eqnarray}
where we have used the analogous equation to Eq.~(\ref{eq:sigmafixed}), valid in the quadratic approximation.
In the quadratic approximation, the expressions in both Eqs.~(\ref{eq:sig1}) and (\ref{eq:sig2}) do not depend on the exact value of $\alpha_s$ used, but it
is most reasonable to use the value of $\alpha_s=\alpha_{\rm WA}$.
Thus, $\Sigma_1$ is the uncertainty-squared in the Higgs boson cross section due to the PDFs with the strong coupling fixed at $\alpha_{\rm WA}$, and $\Sigma_2$ is just the square of the difference
in the best-fit cross sections with the strong coupling fixed at $\alpha_{\rm WA}$ and at
$\alpha_{\rm WA}+\delta{\alpha}_{\rm WA}$.  This is exactly the standard prescription for obtaining the combined
PDF$+\alpha_s$ errors used in the Hessian method.

\section*{Appendix 2: Benchmark calculations of Higgs boson cross sections
\label{sec:APPENDIX3}}

For completeness, we show some benchmark calculations of Higgs boson cross sections
at the LHC
in the tables below. Table~\ref{aptab1} shows a benchmarking comparison of the
inclusive cross sections of
the SM Higgs boson production at the LHC through the gluon fusion subprocess at the LO,
NLO, and NNLO in QCD from different numerical programs. The mass of the Higgs boson, as
well as the renormalization and factorization scales, are set to 125 GeV. We use the
best-fit PDF from the CT10H NNLO analysis for these calculations,
unless otherwise specified. For both MCFM 6.3~\cite{mcfm} and HNNLO 1.3~\cite{hnnlo1,hnnlo2},
the results were calculated in the heavy-quark effective theory (HQET) with infinite top quark mass (scheme A).
In Table~\ref{aptab1} we also show results from iHixs 1.3~\cite{ihixs} using the same setting.
We observe good agreement between the different programs within the numerical accuracy.
For comparison, we also give the results from iHixs with different settings in
Table~\ref{aptab2}; including scheme B, using HQET with finite top quark mass effects
through LO, NLO, and NNLO; scheme C, exact calculations with full dependence on the
top and bottom quark masses up to NLO and with NNLO QCD corrections from HQET with finite
top quark mass; and scheme D, that further incorporates the electroweak (EW) and
mixed QCD-EW corrections.

\begin{table}[h!]
  \begin{center}\scalebox{0.8}{
      \begin{tabular}{c|ccc|ccc|ccc}
        \hline \hline
    &\multicolumn{3}{c}{7 TeV}&\multicolumn{3}{c}{8 TeV}&\multicolumn{3}{c}{14 TeV} \\
    \hline
      (pb)  & MCFM & HNNLO & iHixs& MCFM & HNNLO & iHixs& MCFM & HNNLO & iHixs\\
    \hline
    LO & 4.37 & 4.37 &4.38& 5.58 & 5.58 &5.59& 14.41 & 14.41 &14.50 \\
    \hline
    NLO & 9.96 & 9.98 & 9.99& 12.73 & 12.77 &12.77& 33.05 & 33.15 &33.27 \\
    \hline
    NNLO & -- & 13.38 & 13.50& -- & 17.04 &17.23& -- & 44.49 &44.65 \\
    \hline \hline
      \end{tabular}}
  \end{center}
  \caption{\label{aptab1}
  Inclusive cross sections of the SM Higgs boson production at the LHC through gluon
  fusion at the LO, NLO, and NNLO in QCD from different numerical programs.}
\end{table}

\begin{table}[h!]
  \begin{center}\scalebox{0.8}{
      \begin{tabular}{c|cccc|cccc|cccc}
        \hline \hline
    &\multicolumn{4}{c}{7 TeV}&\multicolumn{4}{c}{8 TeV}&\multicolumn{4}{c}{14 TeV} \\
    \hline
       (pb) & A& B& C & D& A& B& C & D& A& B& C & D\\
    \hline
    LO & 4.38 & 4.68 & 4.38 &4.60& 5.59 &5.97 & 5.60 &5.88& 14.50 &15.53 & 14.55 &15.29 \\
    \hline
    NLO & 9.99 &10.66 & 10.21 & 10.72&12.77& 13.63 & 13.06 &13.72& 33.27&35.58 & 34.16 &35.85 \\
    \hline
    NNLO & 13.50&14.42& 13.97 & 14.65& 17.23&18.40 & 17.83 &18.71&44.65& 47.69 & 46.25 &48.51 \\
    \hline \hline
      \end{tabular}}
  \end{center}
  \caption{\label{aptab2}
  Inclusive cross sections of the SM
   Higgs boson production at the LHC through gluon
  fusion at the LO, NLO, and NNLO with different settings from the program iHixs.}
\end{table}

In Table~\ref{aptab3} we compare the predictions for the inclusive cross sections of the
SM Higgs boson production at the LHC through gluon fusion channel,
calculated with the central-fit PDFs of CT10 and CT10H
NNLO, using HNNLO with the scales set to $M_H$ or $M_H/2$. It can be seen that the
central values increase by only about 0.2\% when comparing CT10H to CT10 predictions.

\begin{table}[h!]
  \begin{center}\scalebox{0.8}{
      \begin{tabular}{c|cccc|cccc|cccc}
        \hline \hline
    &\multicolumn{4}{c}{7 TeV}&\multicolumn{4}{c}{8 TeV}&\multicolumn{4}{c}{14 TeV} \\
    \hline
       (pb) &\multicolumn{2}{c} {CT10} & \multicolumn{2}{c} {CT10H} & \multicolumn{2}{c} {CT10} & \multicolumn{2}{c} {CT10H}
       & \multicolumn{2}{c} {CT10} & \multicolumn{2}{c} {CT10H}\\
    \hline
        & $M_H$ &  $M_H/2$ &  $M_H$ &  $M_H/2$ & $M_H$ &  $M_H/2$ & $M_H$ &  $M_H/2$ & $M_H$ &  $M_H/2$ & $M_H$ &  $M_H/2$\\
    \hline
    LO &4.38&5.30&4.39&5.31&5.59&6.70&5.60&6.71&14.44&16.54&14.47&16.57\\
    \hline
    NLO &9.95&11.88&9.96&11.90&12.71&15.08&12.72&15.13&33.00&38.47&33.03&38.53\\
    \hline
    NNLO  &13.36&14.75&13.38&14.78&17.02&18.91&17.04&18.94&44.41&47.72&44.49&47.79\\
    \hline \hline
      \end{tabular}}
  \end{center}
  \caption{\label{aptab3}
  Inclusive cross sections of the SM Higgs boson production at the LHC through gluon
  fusion at the LO, NLO, and NNLO in QCD with scales equal to $M_H$ or $M_H/2$ and the
  best-fit PDFs of CT10 or CT10H NNLO fits, using the program HNNLO.}
\end{table}

Finally, in Table~\ref{aptab4} we compare the predictions for the production cross sections of
the SM Higgs boson through the gluon fusion
and the vector boson fusion (VBF) processes at the LHC, with 7, 8 and 14 TeV center-of-mass energies.
The VBF cross sections were calculated up to NLO using the
VBFNLO-2.6.1~ \cite{vbfnlo} code, with both the renormalization and the
factorization scales set to $\mu=M_H = 125\ {\rm GeV}$,
and with all the other default settings,
including a minimal invariant mass cut of $600\ {\rm GeV}$
for the two tagging jets. The jet selection cuts
are $p_T>20\,{\rm GeV}$ and $|y|<4.5$, with the $k_T$ jet
algorithm and a distance parameter $D=0.8$. (Neither NLO electroweak
correction nor third jet veto is applied in the calculation.)
In the table, the combined PDF and $\alpha_s$ uncertainties and the
PDF-only uncertainties (inside the parentheses),
evaluated at the $90\%$ C.L., have been calculated by the Hessian method with the
CT10 and CT10H NNLO error PDFs.

\begin{table}
\begin{center}
\begin{tabular}{l|l|ccc}
\hline \hline
 (pb)&  & 7 TeV & 8 TeV & 14 TeV \\
\hline
 \multirow{2}{*}{gluon fusion} & CT10H &
 $13.4^{+4.7(3.0)\%}_{-4.6(3.0)\%}$ &
 $17.0^{+4.8(3.2)\%}_{-4.7(3.1)\%}$ &
 $44.5^{+5.4(4.3)\%}_{-5.0(3.6)\%}$ \\
        \cline{2-5}
 & CT10 &
 $13.4^{+4.7(2.8)\%}_{-5.0(3.1)\%}$ &
 $17.0^{+4.6(2.8)\%}_{-5.0(3.5)\%}$ &
 $44.4^{+4.6(3.1)\%}_{-5.4(4.2)\%}$ \\
\hline
 \multirow{2}{*}{VBF} & CT10H &
 $0.326^{+3.6(3.5)\%}_{-3.7(3.7)\%}$ &
 $0.455^{+3.1(3.1)\%}_{-3.6(3.6)\%}$ &
 $1.454^{+2.6(2.6)\%}_{-4.1(4.0)\%}$ \\
        \cline{2-5}
 & CT10 &
 $0.326^{+4.3(4.3)\%}_{-2.9(2.9)\%}$ &
 $0.454^{+3.9(3.9)\%}_{-2.6(2.6)\%}$ &
 $1.444^{+3.6(3.6)\%}_{-2.6(2.6)\%}$ \\
\hline
\hline
\end{tabular}
\end{center}
\vspace{-2ex}
\caption{\label{aptab4}
Inclusive cross sections of
the SM Higgs boson production
through the gluon fusion
and the VBF processes at the LHC.
The combined PDF and $\alpha_s$ uncertainties
and he PDF-only  uncertainties (inside the parentheses),
at the 90\% C.L.,
have been calculated by the Hessian method with the
CT10 and CT10H NNLO error PDFs.
The errors are expressed as the percentage of the central value.}
\end{table}

\section*{Acknowledgments}
We thank H.-L. Lai and M. Guzzi for useful discussions. T.J. Hou thanks the hospitality of
 the Michigan State University where part of his work was done.
 This work was supported in part by the U.S. National
 Science Foundation under Grant No. PHY-0855561;
by the U.S. DOE Early Career Research Award
DE-SC0003870 and the Lightner Sams Foundation; and by the
 National Natural Science Foundation of China
 under the Grant No. 11165014.

\end{document}